\documentclass{ieeeaccess}
\usepackage{cite}
\usepackage{amsmath,amssymb,amsfonts}
\usepackage{algorithmic}
\usepackage{graphicx}
\usepackage{textcomp}
\usepackage{subcaption}
\usepackage{bm}
\usepackage{multirow}
\usepackage{booktabs}
\usepackage{lscape}       
\usepackage{tabularx}     
\usepackage{booktabs}     
\usepackage{float}
\usepackage{adjustbox}
\usepackage{graphicx} 
\usepackage{colortbl} 
\usepackage[table]{xcolor}
\usepackage{rotating}
\usepackage{adjustbox}
\usepackage{caption}
\usepackage{pdflscape}  
\usepackage{longtable}   
\usepackage{url}
\usepackage{placeins}
\usepackage{graphicx} 

\def\BibTeX{{\rm B\kern-.05em{\sc i\kern-.025em b}\kern-.08em
    T\kern-.1667em\lower.7ex\hbox{E}\kern-.125emX}}

\begin{document}
\history{Date of publication xxxx 00, 0000, date of current version xxxx 00, 0000.}
\doi{10.1109/ACCESS.2024.0429000}

\title{Evaluation of Objective Image Quality Metrics for High-Fidelity Image Compression}
\author{\uppercase{Shima Mohammadi}\authorrefmark{1} \IEEEmembership{Graduate Student Member, IEEE}, 
\uppercase{Mohsen Jenadeleh}\authorrefmark{2} \IEEEmembership{Member, IEEE},
{\uppercase{Jon Sneyers}}\authorrefmark{3}, 
{\uppercase{Dietmar~Saupe}}\authorrefmark{2}, 
{\uppercase{João Ascenso}}\authorrefmark{1} \IEEEmembership{Senior Member, IEEE}}

\address[1]{Instituto de Telecomunicações/Instituto Superior Técnico, University of Lisbon, 1049-001 Lisbon, Portugal}
\address[2]{University of Konstanz, Germany}
\address[3]{Cloudinary, Belgium}

\tfootnote{The IST affiliated authors were supported by FCT/MECI through national funds and when applicable co-funded EU funds under UID/50008: Instituto de Telecomunicações. Mohsen Jenadeleh was funded by the Deutsche Forschungsgemeinschaft (DFG, German Research Foundation), Project ID 496858717. Dietmar Saupe was funded by the DFG Project ID 251654672 -- TRR 161.}

\markboth
{Shima Mohammadi \headeretal: Evaluation of Objective Image Quality Metrics for High-Fidelity Image Compression}
{Shima Mohammadi \headeretal: Evaluation of Objective Image Quality Metrics for High-Fidelity Image Compression}

\corresp{Corresponding author: Shima Mohammadi (e-mail: shima.mohammadi@lx.it.pt).}

\begin{abstract}
Nowadays, image compression solutions are increasingly designed to operate within high-fidelity quality ranges, where preserving even the most subtle details of the original image is essential. In this context, the ability to detect and quantify subtle compression artifacts becomes critically important, as even slight degradations can impact perceptual quality in professional or quality sensitive applications, such as digital archiving, professional editing and web delivery. However, the performance of current objective image quality assessment metrics in this range has not been thoroughly investigated. In particular, it is not well understood how reliably these metrics estimate distortions at or below the threshold of Just Noticeable Difference (JND). This study directly addresses this issue by proposing evaluation methodologies for assessing the performance of objective quality metrics and performing a comprehensive evaluation using the JPEG AIC-3 dataset which is designed for high-fidelity image compression. Beyond conventional criteria, the study introduces Z-RMSE to incorporate subjective score uncertainty and applies novel statistical tests to assess significant differences between metrics. The analysis spans the full JPEG AIC-3 range and its high- and medium-fidelity subsets, examines the impact of cropping in subjective tests, and a public dataset with benchmarks and evaluation tools is released to support further research.
\end{abstract}


\titlepgskip=-21pt

\maketitle
\section{Introduction}
Image compression is essential for minimizing the storage and transmission costs associated with visual data. However, lossy compression techniques inherently introduce distortions that can degrade the quality of the reconstructed images. While the severity of these distortions is largely determined by the bitrate, perceptual factors also play a significant role. Human visual perception is not uniformly sensitive to all distortions; instead, the visibility of compression artifacts is influenced by the content of the image. Specifically, distortions in highly textured or complex regions are often masked, whereas those in smooth or uniform areas tend to be more noticeable.

In recent years, there has been a growing focus on high-quality or visually lossless image compression, where artifacts are either minimally noticeable or imperceptible. This trend is particularly relevant in domains where high visual fidelity is essential, such as professional photography and cinematography, digital archiving, cultural heritage preservation, printing and publishing, medical imaging as well as satellite imaging and remote sensing. Nowadays, user expectations for visual quality have increased significantly since images are frequently enlarged (zooming in on details) or viewed on high-resolution displays. As compression algorithms increasingly prioritize the minimization of perceptual distortion, assessing the visual impact of subtle artifacts becomes more challenging. In this context, both subjective evaluation methods and objective image quality assessment (IQA) metrics are essential for guiding the development, assessment, and quality control of compression algorithms that aim to preserve high perceptual fidelity.

\begin{figure}
    \centering
    \includegraphics[width=1\linewidth]{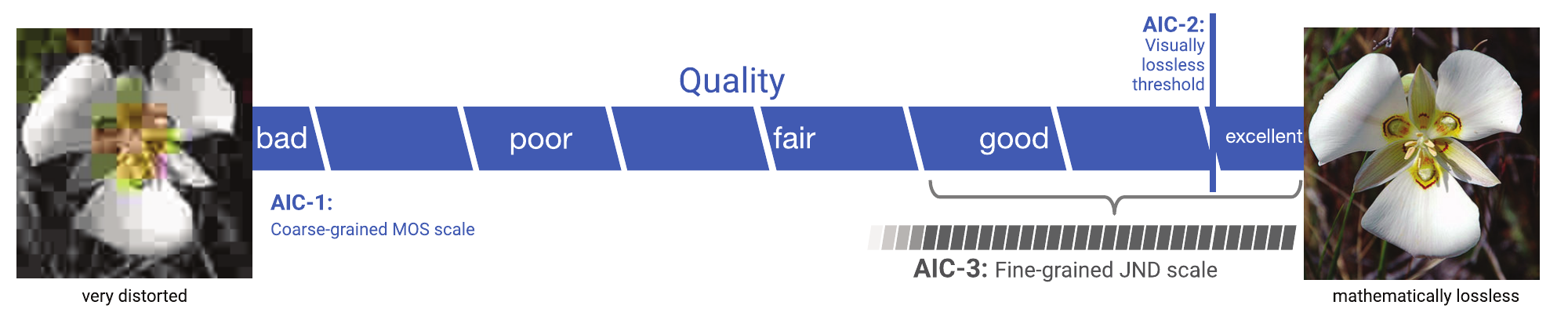}
    \caption{JPEG AIC Quality Range \cite{usecases2025}}
    \label{fig:JPEG_AIC}
\end{figure}

Organizations such as the International Telecommunication Union (ITU) have established standardized procedures for conducting subjective image quality tests, including ITU-T Recommendation BT.500 \cite{BT.500} and ITU-R Recommendation P.910 \cite{P.910}. These methods are widely used in many fields such as image compression, super-resolution, and denoising. However, they often lack the granularity required to distinguish between high-quality outputs generated by recent image compression solutions. This limitation has motivated the development of JPEG AIC-3, an initiative that builds on earlier efforts, AIC-1 \cite{aic1} and AIC-2 \cite{aic2}, to establish best practices for subjective image quality assessments in the range from good quality to mathematically lossless, as illustrated in Figure~\ref{fig:JPEG_AIC}.

The JPEG AIC-3 standard (ISO/IEC 29170-3)~\cite{ISO29170-32025} that is being developed aims to provide a subjective quality assessment methodology for precise subjective image quality assessment in this specific quality range; it leverages triplet-based image comparisons and artifact boosting techniques such as zooming, flickering, and artifact amplification~\cite{men2021subjective, jenadeleh2025, testolina2025fine,jenadeleh2025fine}. In this approach, human observers compare image pairs relative to a reference image, enabling more precise differentiation based on perceived quality. This method enhances sensitivity to subtle compression artifacts, particularly close to the just noticeable difference (JND) threshold, where distortions begin to be perceptible to typical viewers.

Due to the cost and complexity of subjective assessment campaigns, objective IQA metrics that estimate perceptual quality are essential for practical and scalable evaluations. These metrics range from traditional measures such as Peak Signal-to-Noise Ratio (PSNR) and Structural Similarity Index (SSIM) \cite{wang2004image}, to recent deep learning-based metrics such as Learned Perceptual Image Patch Similarity (LPIPS) \cite{zhang2018LPIPS} and Deep Image Structure and Texture Similarity (DISTS) \cite{ding2020image}. However, existing metrics are typically calibrated (or trained) against subjective scores obtained from conventional double (or single) stimulus methodologies such as those outlined in BT.500, for a wide range of quality levels, and thus inherit some limitations. In particular, they tend to saturate in the high-quality range, making them less sensitive to subtle but perceptually significant differences introduced by near-lossless compression. In response, the AIC initiative has issued a call for proposals for metrics, AIC-4, specifically designed to evaluate high-fidelity compressed images \cite{cfpJPEGAIC2025}. 

In this context, there is a notable gap in understanding how well state-of-the-art metrics align with scores obtained from subjective quality assessment methodologies that were recently designed in JPEG AIC to estimate quality. Thus, the main objective of this paper is to study the gap by systematically benchmarking popular metrics in the JPEG AIC-3 quality range, especially for the high-fidelity case, and introduce an evaluation strategy that prioritizes perceptual significance. Therefore, considering this context, the main contributions of this work are:

\begin{itemize}
\item A comprehensive evaluation of existing objective image quality metrics in the range from good quality to mathematically lossless compression. The main goal is to assess how well existing metrics correlate with human perception and thus identify those most suitable for evaluating image compression solutions for the entire JPEG AIC-3 range and for relevant subsets (high-fidelity and medium fidelity).
\item The introduction of new evaluation criteria to assess the sensitivity and reliability of these metrics in the high-fidelity range, namely the z-root mean squared error and its relation to the log likelihood ratio. 
\item Statistical tests to evaluate if a quality metric is statistically significantly better than another, i.e., the observed differences are not due to random chance.
\item Application of the statistical tests to assess the impact of cropping on the full-resolution images that is typically done due to constraints imposed by the platform on subjective assessment.
\item The release of a public dataset with subjective scores, comprehensive benchmarks and evaluation tools (available at \url{https://github.com/shimamohammadi/EvaluationMetrics}). 
\end{itemize}

The subjective assessment methodology, objective quality metrics, transformation and all evaluation measures are described in detail to establish a solid foundation for future research, enabling the rigorous development and validation of quality metrics that are truly aligned with human perception considered the aforementioned context. Considering these contributions, the main impact of this work is on the development and optimization of image codecs, which predominantly rely on metrics due to the impracticality of conducting large-scale subjective studies on a regular basis. Consequently, ensuring that objective quality metrics correlate strongly with human perception is essential for making informed, scalable, and cost-effective decisions during codec design and tuning. 

\section{Related Work}
In the past, many related works have focused on the evaluation of objective quality metrics. Athar et al.~\cite{Zhou2019Access} conducted the most extensive benchmarking study to date of IQA methods, including full-reference, fused-reference, and no-reference metrics on several datasets containing distorted images with a single or multiple distortion types. Their results identify top-performing metrics in each category and demonstrate that rank aggregation-based fusion of full-reference metrics can outperform individual metrics, offering a viable alternative to subjective ratings, especially for large-scale quality assessment datasets where human evaluation is impractical.

Testolina et al.~\cite{Testolina2021a} conducted an extensive performance evaluation of full-reference objective image quality metrics across both conventional (e.g. JPEG, HEVC) and learning-based compression methods. Their results show that metrics with high correlation to subjective quality in traditional codecs often fail to generalize to artifacts produced by deep learning-based codecs, emphasizing the need for more perceptually aligned metrics in recent compression pipelines. 

In a related study, Testolina et al.~\cite{michela2024benchmarking} focused on the high-quality compression scenario, using five source images and five compression methods, including modern codecs like AVIF and JPEG XL. The Double Stimulus Continuous Quality Scale (DSCQS) methodology was used to collect differential mean opinion scores. Experimental results show that learning-based metrics, particularly DISTS, outperform conventional metrics like PSNR and SSIM in correlating with human perception, though all metrics exhibit room for improvement, and often struggle to discriminate subtle quality differences in this quality range, raising concerns about their reliability in perceptually critical applications. 

Li et al.~\cite{Li2021} confirmed the findings of Testolina et al., showing that conventional metrics often fail to generalize to artifacts produced by learning-based compression methods, especially in regions such as textures or complex structures. To address this, they constructed a benchmark dataset using double stimulus experiments with degradation category ratings that includes both traditional and learning-based codecs. The authors also propose a new full-reference self-attention-based quality prediction metric and show through experiments that it better reflects subjective judgments, supporting the need for specialized metrics tailored to learning-based image compression.

Hammou et al.~\cite{hammou2025} proposed an evaluation procedure that recreates classic psychophysical experiments to assess full-reference quality metrics based on key characteristics of human vision, including contrast sensitivity, contrast masking, and contrast matching. These tests provide deeper insight into how well a metric aligns with low level human vision, which is particularly important for evaluating high-fidelity image compression. The evaluation on widely used quality metrics with this framework uncovers sharp differences: deep learning-based metrics like LPIPS and DISTS, as well as relatively simple metrics like SSIM and MS-SSIM, capture contrast masking well, whereas VMAF performs worse in this regard. Additionally, SSIM tends to overemphasize differences in high-frequency regions, a shortcoming which is well addressed by MS-SSIM. However, it is important to note that good performance in these low-level human vision tests may not be a necessary or sufficient condition for a metric to perform well across all application scenarios.

Moreover, Ullah et al.~\cite{ullah2023subjective} performed a flickering subjective test to evaluate objective image quality metrics in visually lossless image compression range with a dataset of images compressed with the JPEG 1 standard. The authors use subjective flicker tests and full-reference image quality metrics to identify the compression factor that ensures ``visually lossless'' JPEG 1 compression. The findings correspond to concrete numerical thresholds to achieve imperceptible quality loss, bridging the gap between time-consuming subjective tests and fast, scalable objective assessment.

Cheon et al.~\cite{AmbiguityIQA2016} proposed a new evaluation framework that considers both accuracy and ambiguity in assessing image quality metrics. This work introduced the concept of an ambiguity interval, the range within which different metric scores may be perceptually indistinguishable to human viewers. Their experiments reveal that a metric with the highest accuracy does not necessarily perform best when ambiguity is considered since it may lead to a metric with low reliability, sensitivity, or distinguishability.

Sneyers et al. \cite{ssimulacra2} introduced a novel subjective testing methodology that combines pairwise comparison and double stimulus impairment scale protocols, specifically tailored for evaluating image quality in the high-fidelity compression range. As part of this work, the Cloudinary Image Dataset’22 dataset was created, which is the result of a large-scale subjective experiment to benchmark and develop perceptually aligned image quality metrics.

While these studies have significantly advanced the understanding of perceptual quality assessment, most rely on datasets collected using double stimulus methodologies and do not fully capture perceived impairments in JND units from good quality to mathematically lossless compression. Our study helps bridge this gap by systematically benchmarking widely used metrics and proposing evaluation strategies that prioritize perceptual significance over purely numerical accuracy, especially within the AIC-3 spectrum. Moreover, compared to \cite{jenadeleh2025}, this paper evaluates a broader range of quality metrics, introduces additional evaluation measures, and proposes a novel criterion to compare metric performance, with a specific focus on sensitivity and reliability in the high-fidelity compression range. Furthermore, statistical tests are introduced and the analysis is extended to include evaluations across targeted quality sub-ranges and considers both the cropped versions used in subjective tests and the full-resolution test images.

\section{JND-Based Subjective Assessment Methodologies}\label{sec:rev_subj}
JND-based assessment can be approached from two ways. First, it can refer to the estimation of the JND threshold, i.e., the minimal level of distortion that is detected by a random observer with a probability of 50\%. Importantly, the JND is context-dependent and thus perception by human observers varies depending on the conditions under which images are evaluated. Factors such as display size, viewing distance, ambient lighting, and the use of single- or double-stimulus methodologies play a major role. Thus, there is no universally accepted full definition of the JND threshold. Whenever JND threshold values are reported, the corresponding context conditions should be stated. Secondly, JND-based assessment can refer to the estimation of perceived quality or impairment on a continuous scale having units that are based on just noticeable difference. This JND-scale implies that any two distorted stimuli with a difference of 1 JND unit on the scale will be discerned by a random observer with 50\% probability. 

\subsection{JND threshold estimation}
Regarding JND threshold estimation, several methods exist to identify the distortion level (or encoding parameter) that yields a compressed image at the JND threshold. In controlled laboratory studies \cite{lin2015experimental, jin2016statistical, wang2016mcl, videoset}, binary search algorithms were employed to compare original and distorted stimuli to estimate observer-specific thresholds. Crowdsourced JND evaluations, employing interactive sliders with flicker testing, offer a scalable alternative \cite{lin2022large,lin2020subjective}. In \cite{jenadeleh2024crowdsourced}, a state-of-the-art adaptive psychometric testing methods, QUEST+~\cite{watson2017questplus}, has been adapted for crowdsourced estimation of the JND for compressed videos by treating the entire population as a collective observer. This method randomly selects individual observers for each paired comparison between a reference stimulus and its compressed versions, enabling efficient and reliable estimation of the population-level JND threshold.

A flicker test methodology, which was standardized as ISO/IEC 29170-2, also known as JPEG AIC-2 \cite{aic2} can be used to reliably estimate the threshold for visually lossless compression. In this test, a reference (uncompressed) image is presented alongside an image composed of alternating views of the reference and a distorted version. Often this sequential alternation of the reference and compressed images is done at a high frequency, such as 8\,Hz. The test subjects have the task of identifying the image that does not flicker. This methodology allows to classify compressed images as either visually lossless or visually lossy, and thus defines the boundary between lossy and lossless compression as the JND threshold under flicker test conditions. 

\subsection{Quality assessment on the JND-scale}
The procedures mentioned in the previous subsection are applied to estimate the JND threshold by comparing a reference stimulus with its compressed versions. In the context of image compression, where perceptually lossless or near-lossless compression is often desired, it becomes essential to employ highly sensitive testing procedures and to obtain measurements on a continuous scale, for a wide range of distorted images, including those compressed at very high quality factors. A subjective assessment methodology that provides such scale values from comparisons, was introduced by Thurstone \cite{TM} and first applied for assessment of distortion from compression artefacts in \cite{watson2001measurement}. This framework assumes a model for pair comparisons of compressed images each of which has a latent perceptual quality score, given as an independent random variable with a normal distribution of fixed variance. Then differences between these scores determine probabilities that the image with larger distortion is correctly identified in a 2-alternative forced choice comparison. 

For a particular choice of variance a difference of 1 unit corresponds to a 75\% proportion of correct responses, i.e., there is a 50\% probability that a typical observer will notice the difference in quality between two images. The scores are adjusted so that the quality of the reference image is set to zero. This way, the quality scale is continuous and gives the perceived impairments in JND units, ranging from zero to infinity. Several works have applied these models \cite{watson2001measurement,testolina2023jpeg,men2021subjective,perez2017practical,zerman2019analysing}.

\subsection{Fine-grained quality assessment with boosted triplet comparisons}
Visually comparing very small image impairments at or below the JND threshold is challenging, and the resulting JND scale values can be expected to be noisy. To address this particular difficulty at high visual quality levels, the upcoming JPEG AIC-3 standard \cite{ISO29170-32025} was introduced.

The JPEG AIC-3 protocol employs boosted triplet comparisons (BTC) to make subtle distortions more noticeable, offering a reliable procedure for evaluating compression performance when visual distortions are minimal. The boosting techniques defined include three types of processing: 1) zooming: each image is enlarged two times; 2) artifact amplification: the pixel-wise numerical difference between the reference and distorted versions is doubled in every color channel; and 3) flicker: the reference and distorted frames alternate at 10\,Hz, with each shown for 100\,ms. For every compressed image, boosted variants are obtained by applying the zoom and artifact amplification steps, while the flicker sequence is generated in real time when subjects view the triplets. In the boosted-triplet comparison (BTC) setup, the subjects pick the image of the pair that flickers most strongly or choose ``Not sure'' when no clear difference is perceived. It has been shown that offering a ``Not sure'' response option reduces mental load while preserving the homogeneity of the psychometric function \cite{jenadeleh2023relaxed}. BTC is complemented with plain-triplet comparison (PTC), where no boosting technique is applied. For PTC, users switch between the compressed and reference images with a toggle button, each trial demands at least one toggle, and interaction is restricted to a maximum of two toggles per second. In PTC, subjects compare two compressed images to the original and select the one that appeared more degraded. In the BTC experiment, the full range of distortion levels plus the reference is evaluated, whereas the PTC variant samples every second level, assessing a lower number of distortion levels. In both BTC and PTC, responses are collected using a pairwise comparison format, with the options being ``Left'', ``Right'', or ``Not Sure''. For the data analysis, the responses of type ``Not Sure'' are split equally between ``Left'' and ``Right''.

\subsection{Scale Reconstruction}\label{sec:aic3_processing}
For the scale reconstruction, a joint functional model for PTC and BTC can be applied as defined in the JPEG AIC-3 standard \cite{ISO29170-32025}. Per source image and codec combination, the perceived distortion for the unboosted stimuli is modeled by an exponential rate–distortion curve, $d(r)=\alpha e^{-\beta r}$, having parameters $\alpha$ and $\beta$. Then, the perceived distortion for the boosted stimuli is given by the quadratic mapping  $t(d)=\gamma_{1}d+\gamma_{2}d^{2}$, that converts the unboosted scale $d(r)$ into its boosted counterpart $t\!\bigl(d(r)\bigr)$. Thus, per source image and codec combination, four parameters need to be estimated, $\alpha$, $\beta$, $\gamma_{1}$, and $\gamma_{2}$. This analysis is carried out by maximum likelihood estimation, using the Thurstonian Case V model~\cite{TM} to define the likelihoods. Finally, bootstrapping is applied 1,000 times to estimate the distribution of quality estimates, providing the mean ($\mu$) and standard deviation ($\sigma$) for each image.

\section{JPEG AIC-3 dataset}\label{sec:aic3_dataset}
The JPEG AIC-3 dataset includes five high-quality source images which are shown in Figure~\ref{fig:dataset}. All images were cropped to a resolution of $620\times800$ pixels prior to testing. In the first experiment \cite{testolina2025fine}, each image was compressed using five lossy image codecs, AVIF, JPEG, JPEG 2000, JPEG XL and VVC Intra, at ten different quality levels. The comparisons included: i) same-codec comparisons between images compressed with the same codec at different distortion levels, ii) cross-codec comparisons across codecs to align quality scales, iii) trap comparisons with the most distorted image (level 10) with its source to detect unreliable subjects. The web interfaces developed by JPEG~AIC-3 for BTC and PTC were used to obtain the dataset. Participants were recruited through Amazon Mechanical Turk (MTurk) platform for both BTC and PTC, experiments performed separately. Participants could complete a maximum of two different batches, with questions presented in a randomized order for each individual. This dataset was recently extended in \cite{jenadeleh2025} to include a comprehensive subjective visual quality assessment of JPEG~AI image compression solution also using the JPEG AIC-3 methodology. A limited number of objective quality metrics was evaluated on this dataset just using SROCC and PLCC over the entire dataset. Together, a total of 300 compressed images with scale values and standard deviations have resulted, and these are used in the evaluation performed in Section \ref{sec:exp-results}.

\newcommand{\putimg}{\includegraphics[height=4.0cm]}
\newcommand{\putimgt}{\includegraphics[height=2.2cm]}
\newcommand{\commentout}[1]{}
\begin{figure*}[!htbp]
    \centering
    \setlength{\tabcolsep}{3pt}
    \begin{tabular}{ccccc}
    \putimgt{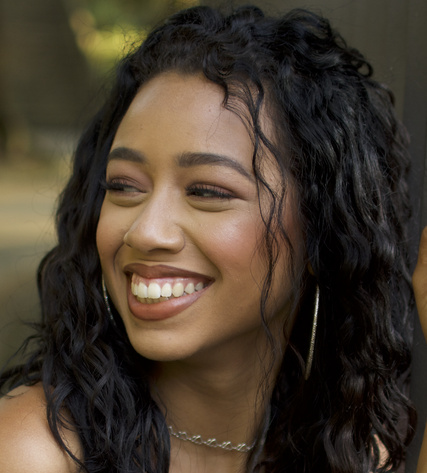}
    & \putimgt{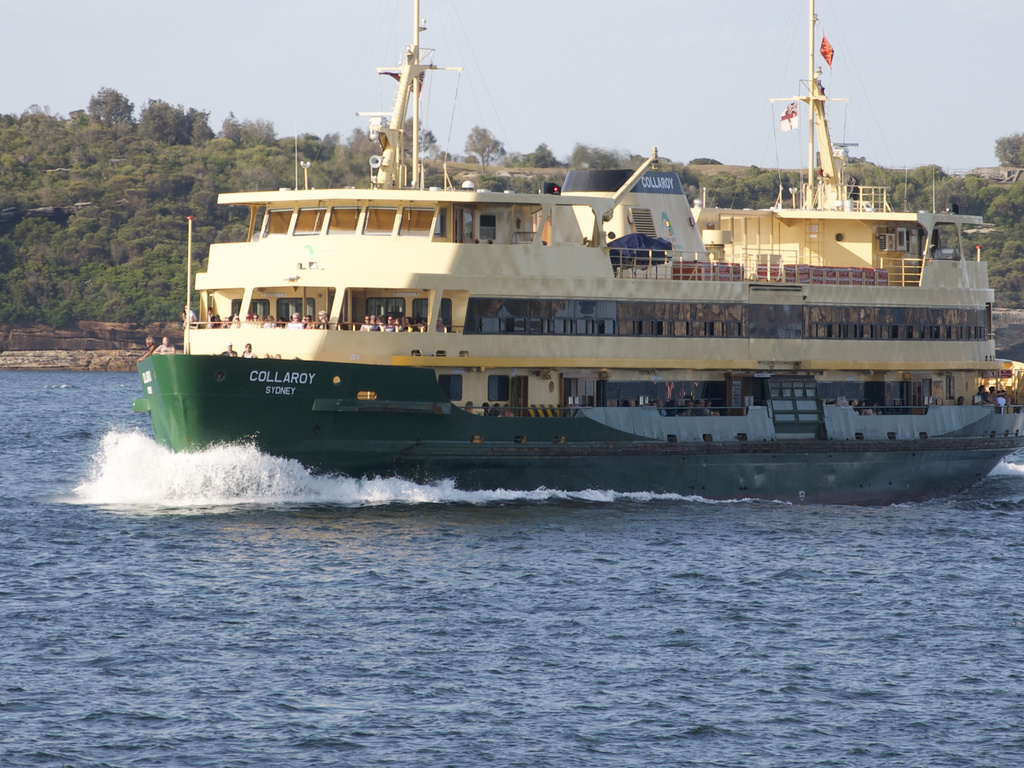}
    & \putimgt{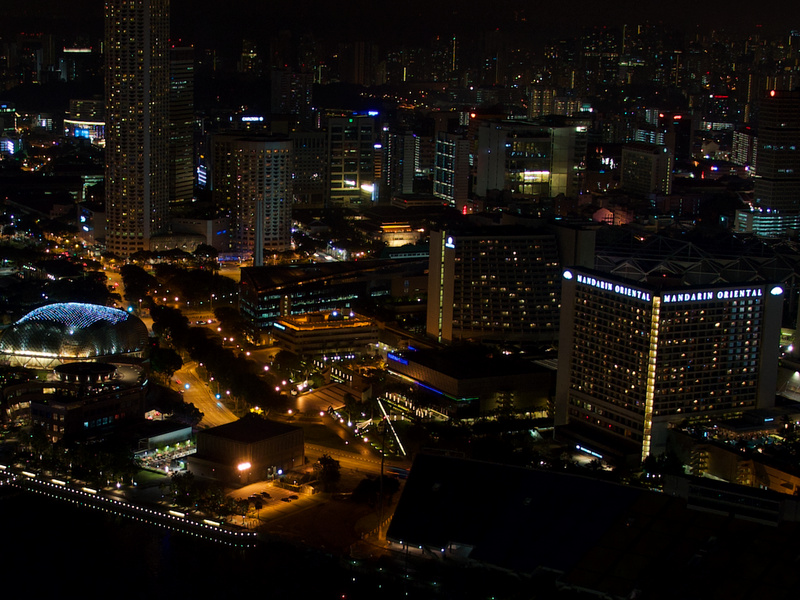}
    & \putimgt{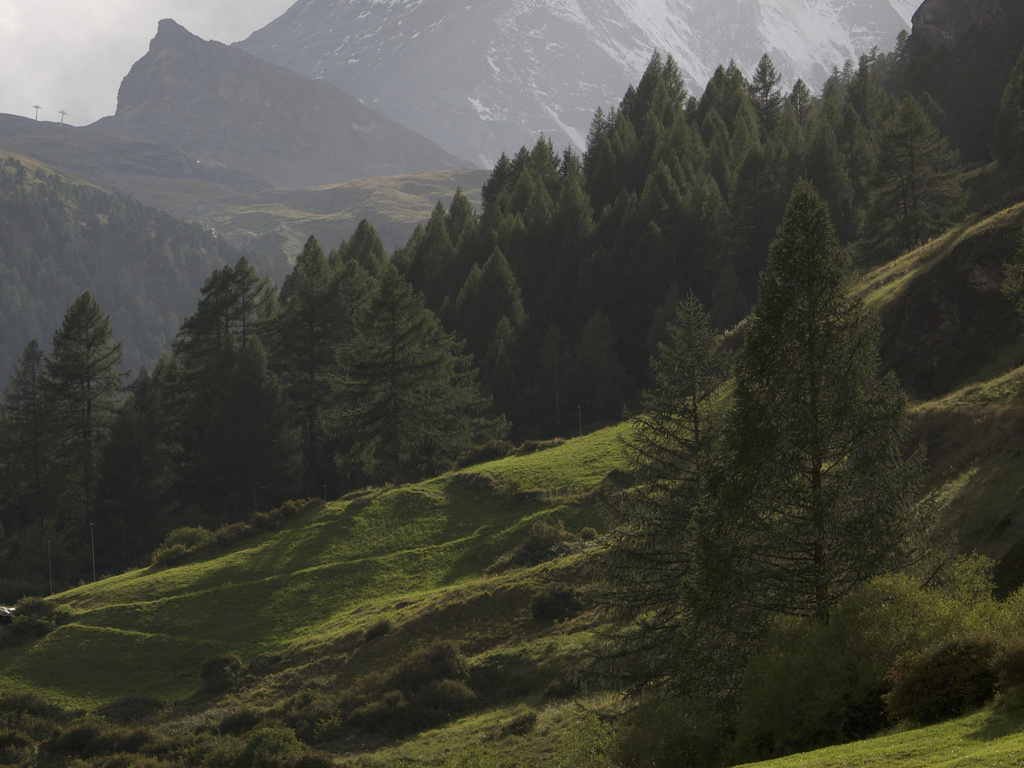}
    & \putimgt{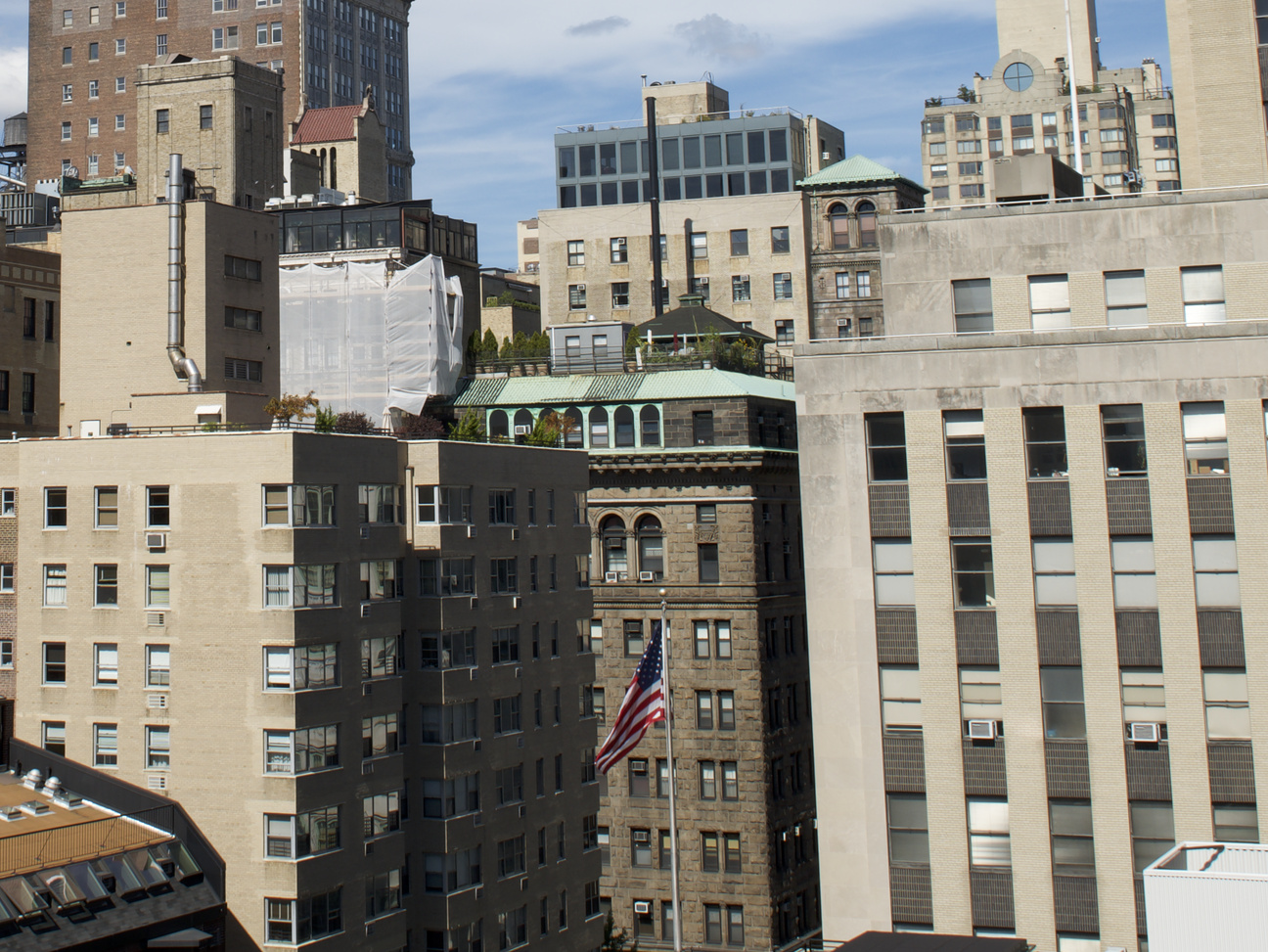}
    \\
    \putimg{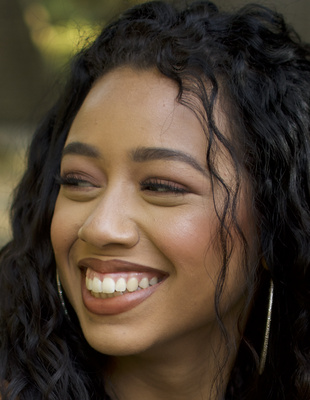}
    & \putimg{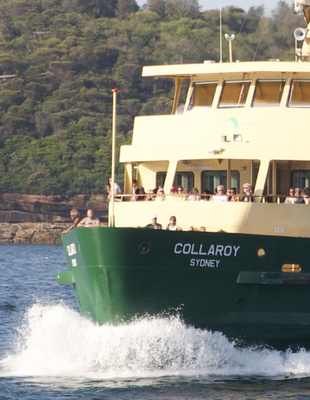}
    & \putimg{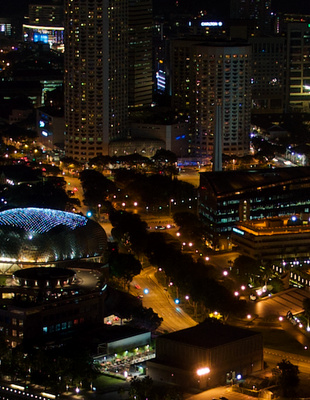}
    & \putimg{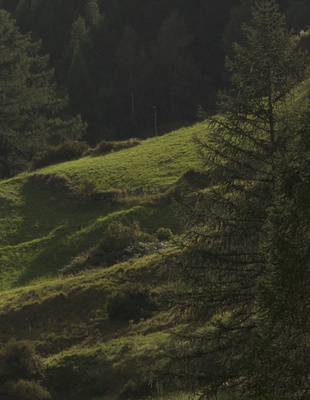}
    & \putimg{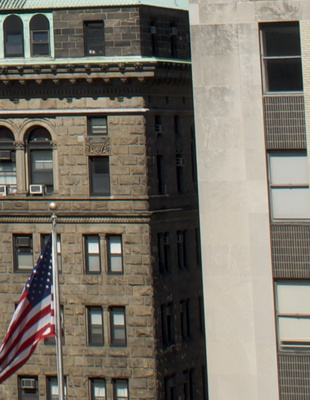}
    \\

     SRC 00002 & SRC 00006 & SRC 00007  & SRC 00009  & SRC 00010
    \end{tabular}%
    \caption{Full resolution images and the crops used in the subjective assessment experiment.}
     \label{fig:dataset}
\end{figure*}
\section{Objective Quality Metrics}\label{obj_quality_metrics}
The main goal of an objective quality assessment is to correlate strongly with human perception. In the case of image compression, such metrics are essential for making informed, scalable, and cost-effective decisions during codec design and online optimization. Nowadays, full-reference image quality metrics can be classified into three approaches:

\begin{itemize}

\item \textbf{Conventional algorithmic approaches:} These methods apply hand‑crafted mathematical models to quantify pixel‑wise or structural discrepancies between the reference and the test image. Classic examples include mean‑squared error (MSE) and its logarithmic version, the \emph{peak signal‑to‑noise ratio} (PSNR), the \emph{structural similarity index} (SSIM)\cite{wang2004image}, its multi‑scale extension (MS‑SSIM) \cite{wang2003multiscale}, the \emph{feature similarity index} (FSIM)\cite{zhang2011fsim}, the \emph{information fidelity criterion} (IFC), and many others. Because they require only a few simple operations (differences, convolutions, or frequency transforms) they are fast, interpretable, and relatively easy to integrate into coding pipelines. However, due to the use of low‑level signal statistics, they may have difficulty in predicting human opinion when there are high‑level semantics or complex mixed distortions.

\item \textbf{Learning‑based approaches:} In the past decade, deep neural networks have been trained end‑to‑end to predict subjective scores. These models represent the reference and distorted images (often patch‑wise) into a learned feature (or latent) space and often use regression models to estimate quality. Representative metrics include \emph{Learned Perceptual Image Patch Similarity} (LPIPS)\cite{zhang2018LPIPS}, \emph{Deep Image Structure and Texture Similarity} (DISTS)\cite{ding2020image}, \emph{PieAPP}, and deep Wasserstein Distance (deepWSD) \cite{DeepWSD}. These metrics are trained on hundreds of thousands of images and corresponding subjective scores and are able to capture complex artifacts such as texture misalignment, global contrast shifts, and semantic inconsistencies that may not be easy to predict with conventional metrics. Their drawbacks are the need for large, curated datasets of subjective scores, higher computational cost, and possibly performance when evaluated outside the distortion domain seen during training (overfitting).

\item \textbf{Multi‑method fusion approaches:} Rather than betting on a single descriptor, fusion schemes aggregate a rich set of heterogeneous quality features and learn an ensemble mapping to mean opinion scores. The most popular example is \emph{Video Multi‑method Assessment Fusion} (VMAF)\cite{vmaf} developed by Netflix, which combines PSNR, SSIM, detail‑loss metrics, and temporal information using support‑vector regression. Some variants extend the idea with resolution‑aware or content‑adaptive features. By blending complementary cues, fusion approaches achieve state‑of‑the‑art correlation with subjective judgement across codecs, contents, and display resolutions. Their main costs are increased complexity, the need for periodic re‑training as encoders evolve, and the lack of interpretation ability of the final regression model.
\end{itemize}

In practice, the choice of a quality metric involves a trade‑off between prediction accuracy, computational footprint, and the types of distortions most relevant to the target application (e.g. compression, super‑resolution). The quality metrics listed in Table~\ref{tbl-quality-metrics} represent a diverse set of approaches, ranging from conventional signal-based measures to modern learning-based models.

\begin{itemize}
    \item \textbf{PSNRY}: Compares each pixel of a reference image with the corresponding pixel in the decoded (or compressed) image. In this metric, all pixels contribute equally to the overall quality score, which is calculated here just for luminance (Y). However, PSNR often does not correlate well with human perception of image quality, especially when distortions are spatially non-uniform.
    \item \textbf{PSNR-HVS}: PNSR that operates on $8\times8$ image blocks using DCT basis functions and accounts for between-coefficient masking and the contrast sensitivity function (CSF) to estimate the maximum imperceptible distortion.
    \item \textbf{SSIM}\cite{wang2004image}: Evaluates the similarity between two images based on three components: luminance, contrast, and structure. These components are combined into a single SSIM score using a weighted formulation. Unlike traditional metrics such as PSNR, SSIM is designed to better reflect human visual perception of image quality.
    \item \textbf{MS-SSIM}\cite{wang2003multiscale}: An extension of SSIM that incorporates variations in image resolution and content by evaluating the structural similarity at multiple scales. This is achieved by iteratively applying a low-pass filter and down-sampling the image by a factor of 2. The original image is considered as the first scale, while the coarsest scale (scale $M$) is obtained after $M{-}1$ iterations. At each scale $j$, contrast and structure comparisons are computed, while luminance comparison is performed only at the coarsest scale. The final MS-SSIM score is obtained by combining the measurements from all scales using a weighted geometric mean
    \item \textbf{IW-SSIM}\cite{iw-ssim}: An extension of SSIM that incorporates the notion of information content by assigning weights to different image regions based on their local structural complexity. Unlike MS-SSIM, which focuses on multiple scales, IW-SSIM emphasizes perceptually important areas by giving lower weights to regions rich in texture and edges, and higher weights to smooth regions. This results in a metric that better correlates with human perception of image quality, especially in images with spatially varying distortions.
    \item \textbf{FSIM}\cite{zhang2011fsim}: Evaluates the similarity between two images based on low-level features, primarily phase congruency and gradient magnitude. Phase congruency captures significant structural information that is invariant to contrast changes, while gradient magnitude reflects local contrast. FSIM combines these complementary features by weighting the similarity scores with phase congruency, resulting in a quality measure that aligns well with human visual perception.
    \item \textbf{FSIMc}: An extension of FSIM that incorporates color information to better assess image quality for color images. FSIMc combines the original FSIM features,phase congruency and gradient magnitude, with chrominance components obtained from the YIQ color space. This integration allows FSIMc to capture both structural and color distortions, providing a more comprehensive and perceptually accurate evaluation of color image quality.
    \item \textbf{GMSD} \cite{xue2013gradient}: Evaluates the similarity between two images by computing a patch-wise gradient magnitude similarity (GMS) map. The standard deviation of each GMS map is used as a pooling strategy to capture local variations in quality degradation, reflecting perceptual differences across the image.
    \item \textbf{UQI}\cite{wang2002universal}: Models image distortion as a combination of loss of correlation, luminance distortion, and contrast distortion between a reference and a distorted image. It computes these components locally and combines them into a single quality score as the product of the three components.
    \item \textbf{NLPD}\cite{laparra2016perceptual}: Operates in the normalized Laplacian pyramid domain. The input images are first decomposed into multiple spatial scales using a Laplacian pyramid, which removes local luminance variations at each scale. Each resulting decomposition is then normalized by a locally estimated amplitude to account for contrast variations. Image quality is assessed by computing the root mean squared error between the corresponding normalized coefficients of the reference and distorted images.
    \item \textbf{VIF}\cite{sheikh2006image}: Quantifies image fidelity by measuring the amount of information in the reference image and evaluating how much of that information can be extracted from the distorted image (mutual information) in the wavelet domain. It also introduces a Human Visual System (HVS) model that incorporates perceptual uncertainty during the information processing stage.
    \item \textbf{VMAF}\cite{vmaf}: Video quality assessment metric developed by Netflix that estimates perceptual quality by combining the outputs of several individual quality metrics using a support vector machine (SVM) regression model. By learning from subjective quality scores, VMAF produces a single quality prediction that aligns closely with human perception. Although originally designed for video content, it can also be effectively applied to still images.
    \item \textbf{VMAF-NEG}\cite{vmaf-neg}: Variant of the VMAF metric designed to prevent quality scores from increasing when the distorted image appears better than the reference due to enhancement artifacts. VMAF-NEG modifies the VMAF fusion model to avoid giving a quality gain to artificially enhanced or over-sharpened content. 
    \item \textbf{HaarPSI}\cite{reisenhofer2018haar}: HaarPSI utilizes the Haar wavelet transform to decompose images into multiple frequency components. It focuses on high-frequency coefficients to assess local similarities and employs low-frequency coefficients to weight the importance of different image regions, modeling human sensitivity with a logistic function.
    \item \textbf{VSI}\cite{zhang2014vsi}: Incorporates visual saliency to better align with human perception. It combines local similarity in gradient magnitude and chrominance with a visual saliency map that emphasizes regions likely to attract human attention. 
    \item \textbf{LPIPS}\cite{zhang2018LPIPS}: Perceptual image quality metric that compares deep features extracted from a pre-trained neural network to measure visual similarity between two images. LPIPS computes the distance between feature maps of the reference and distorted images across multiple layers of a CNN (e.g. AlexNet or VGG), using learned weights to better reflect human perceptual sensitivity. It captures both low-level and high-level differences, providing a similarity score that strongly correlates with subjective assessments.
    \item \textbf{DISTS}\cite{ding2020image}: Deep learning–based image quality metric designed to balance sensitivity to structural distortions with robustness to texture variations. It maps both the reference and distorted images into a feature space using a modified VGG-16 network, and compares them using both spatial averages (for texture) and feature correlations (for structure), producing a perceptually meaningful similarity score.
    \item \textbf{A-DISTS}\cite{ding2021adists}: An extension of the DISTS metric that introduces adaptive weighting between structure and texture similarity components. A-DISTS uses deep features from a pre-trained CNN (similar to DISTS) but dynamically adjusts the relative importance of structural and textural features for each image pair. 
    \item \textbf{PieAPP}\cite{PieAPP}: A deep learning–based perceptual image quality metric trained on pairwise preferences obtained through subjective testing. PieAPP predicts the probability that one distorted image is perceived as higher quality than another; it uses a CNN to extract features from image patches and learns to aggregate them into a global quality score.
    \item \textbf{FLIP}\cite{Andersson2020}: Designed to highlight perceptual differences between two images on a per-pixel basis, closely mimicking how a human would perceive changes when flipping between the images. It generates per-pixel error maps that consider viewing conditions such as display resolution and observer distance, utilizing a perceptually uniform color space to accurately model visual differences. FLIP also places particular emphasis on edge regions, where minor variations are more perceptible.
    \item \textbf{CVVDP}\cite{cvvdp}: A perceptual image quality metric that models chromatic and achromatic contrast sensitivity as well as spatio-temporal sensitivity of the human visual system. It accounts for viewing conditions and display characteristics to predict the visibility of distortions accurately. CVVDP utilizes a novel psychophysical model of chromatic spatiotemporal contrast sensitivity and cross-channel contrast masking, providing perceptually meaningful assessments of visual quality
    \item \textbf{TOPIQ}\cite{chen2023topiq}: Emulates the human visual system by exploiting high-level semantic information to guide the assessment of local distortions. It employs a coarse-to-fine neural network with a cross-scale attention mechanism, where semantic features from higher layers guide the assessment on semantically significant regions in lower layers. 
    \item \textbf{SSIMULACRA2}\cite{ssimulacra2}: Builds on MS-SSIM and operates in a perceptually aligned color space (XYB), applies downscaling in linear RGB, and uses additional error maps to specifically detect compression artifacts like blockiness and blurring. Several measures are calculated from error maps across multiple scales, which are aggregated using both the $L^1$ and $L^4$ norms, with weights tuned using Nelder-Mead simplex search to minimize MSE and to maximize Kendall and Pearson correlation for a training dataset. SSIMULACRA1 is an early version of this metric which works in the CIE Lab color space instead of XYB.   
    \item \textbf{BUTTERAUGLI}\cite{butteraugli}: Inspired by the statistical variations in the location and density of different color receptors and the modeling of ganglion cells, it simulates how humans perceive differences, particularly subtle artifacts in high-quality compressed images. It produces both an overall similarity score and a heatmap that highlights areas with noticeable changes.
    \item \textbf{HDR-VDP-1/2}\cite{hdr-vdp-mantiuk}: visual metrics based on a contrast sensitivity model valid across all luminance conditions, which predicts both visibility and image quality. The proposed visual model serves as the core of a visual difference predictor, estimating the likelihood that an average observer will notice differences between two images. These metrics are specifically tuned to detect near-threshold, just-noticeable differences. 
\end{itemize}

\begin{table}[htb]
\scriptsize
\caption{Configuration options for the selected quality metrics.}
\centering
\renewcommand{\arraystretch}{1.1}
\begin{tabular}{c|c|c|>{\raggedright\arraybackslash}p{3.8cm}}
\hline
\textbf Class & {Metric} & \textbf{Input} & \textbf{Configuration} \\
\hline
\multirow{19}{*}{\rotatebox{90}{\textbf{Conventional}}}
& PSNRY & Y & Default Parameters \\
& PSNR-HVS & Y & Default Parameters\\
& MS-SSIM  & Y & IQA pytorch libary \\
& SSIM  & Y & Scikit-image library \\
& FSIM  & Y & Default Parameters \\
& IW-SSIM & Y &  Default Parameters\\
& GMSD  & Y & Default Parameters \\
& UQI & Y & Default Parameters  \\
& NLPD  & Y & Default Parameters \\
& VIF  & Y & Default Parameters \\
& Haar-PSI  & RGB & Implementation: Python \\
& FSIMc  & RGB & Default Parameters \\
& VSI  & RGB & Default Parameters \\
& CVVDP  & RGB & v0.4.2; Disp: standard Full HD \\
& FLIP  & RGB & Score: Mean \\
& SSIMULACRA2  & RGB & Default Parameters \\
& HDR-VDP-2 Q  &linear RGB & v2.2.2; pixels per degree: 64.27; display: ccfl lcd \\
& HDR-VDP-3 Q  &linear RGB & v3.0.7; Task: detection; BT.709; pixels per degree: 64.27; display: ccfl lcd\\
& BUTTERAUGLI  & RGB & Used the three norms variant \\ \hline
\multirow{4}{*}{\rotatebox{90}{\textbf{Fusion}}}
& VMAF  & YUV & Model \texttt{vmaf\_v0.6.1.json} model; 4:4:4@8bit \\
& VMAF-NEG  & YUV & Model: \texttt{vmaf\_v0.6.1neg}; 4:4:4@8bit \\ \hline
\multirow{5}{*}{\rotatebox{90}{\textbf{Learning}}}
& TOPIQ & RGB & Default Parameters \\
& DISTS  & RGB & Implementation: PyTorch \\
& A-DISTS  & RGB & Default Parameters \\
& LPIPS  & RGB &  Default Parameters\\
& PieAPP  & RGB &  Sampling mode: Dense  \\
\hline
\end{tabular}
\label{tbl-quality-metrics}
\end{table}

Some details about the quality metrics are presented in Table~\ref{tbl-quality-metrics}. Luminance-based metrics such as \mbox{PSNRY}, SSIM, and FSIM require grayscale input. The luminance is computed from RGB using ITU-R BT.709 \cite{iturbt709}, i.e. $Y = 0.2125R + 0.7154G + 0.0721B$. For the HDR-VDP-2/3 quality metrics, the luminance is converted to nits ($cd/m^2$).

\section{Transformation} \label{sec:transformation}
For some of the evaluation criteria that will be introduced later, the metric values need to be transformed to be aligned with the subjective scores. Since human perception of image quality is often non-linear, a four parameter logistic function is selected due to its strict monotonicity and balance between providing an accurate fit and maintaining a reasonable level of complexity,
\begin{equation}
    S_{\text{trans}} = B_2 + \frac{B_1 - B_2}{1 + e^{- \frac{s_{\text{obj}} - B_3}{B_4}}}.
    \label{equ-logistic}
\end{equation}
Here $s_{obj}$ represents the objective score of a quality metric that serve as the predictor, while $s_{\text{trans}}$ represents the transformed value used for evaluation. This model is fitted to the data using non-linear least-squares fitting, which optimized the coefficients $B_1, B_2, B_3,$ and $B_4$ to best match the observed data in terms of RMSE (but not other criteria).

\section{Conventional Evaluation Criteria}\label{sec:conv_eval}
In this section, quality metrics evaluation criteria used in previous state-of-the-art are described. 

\begin{itemize}
\item Root mean square error (RMSE): The square root of the average squared differences between the predicted and actual values. RMSE gives a higher weight to larger errors due to the squaring of the residuals, making it sensitive to outliers. A lower RMSE indicates a better fit of the model to the data, with a value of 0 representing a perfect prediction.
\item Pearson linear correlation coefficient (PLCC): A statistical measure that quantifies the strength and direction of the linear relationship between two variables. It is calculated as the covariance of the two variables divided by the product of their standard deviations. The coefficient ranges from -1 to 1, where 1 indicates a perfect positive linear correlation, -1 indicates a perfect negative linear correlation, and 0 indicates no linear relationship. PLCC assumes that the data is normally distributed and that the relationship between variables is linear. It is widely used in regression analysis and other statistical modeling tasks to assess how well changes in one variable predict changes in another.
\item Spearman Rank Order Correlation Coefficient (SROCC): A non-parametric statistical measure that assesses the strength and direction of a monotonic relationship between two variables based on their ranked values rather than their numerical values. Unlike Pearson’s correlation, Spearman’s correlation does not assume a linear relationship or normally distributed data, making it more robust to outliers and applicable to ordinal data. SROCC is calculated by first converting the original values into ranks and then computing the Pearson correlation coefficient between these ranks. The result ranges from -1 to 1, where 1 indicates a perfect positive monotonic relationship (as one variable increases, the other consistently increases), -1 indicates a perfect negative monotonic relationship (as one variable increases, the other consistently decreases), and 0 suggests no monotonic relationship between the variables.
\item Kendall Tau (KT): A non-parametric statistic used to measure the ordinal association between two variables. It evaluates the strength and direction of a monotonic relationship by comparing the number of concordant and discordant pairs. The coefficient ranges from -1 to 1, where -1 indicates a perfect inverse (negative) monotonic relationship, 1 indicates a perfect direct (positive) monotonic relationship, and 0 implies no monotonic association. It is particularly useful when the data do not necessarily follow a linear pattern and is more robust to outliers than Pearson’s correlation.
\item Perceptually Weighted Rank Correlation (PWRC) is a statistical measure that prioritizes the sorting accuracy (SA) or accurate ranking of high-quality images while minimizing the weight of insensitive ranking errors. Unlike SROCC, which measures overall ranking accuracy, PWRC selectively activates pairwise comparisons where the score difference exceeds a sensory threshold (ST), ignoring imperceptible rank differences. Additionally, it applies non-uniform weighting, assigning higher importance to image pairs based on quality level and rank deviation. By adjusting the perception threshold, PWRC enables sorting accuracy visualization through the SA-ST curve, offering a more perceptually relevant assessment of the metrics under evaluation.
\item Outlier Ratio (OR) expresses the percentage of test items for which the metric departs from the subjective opinion score such that the error can no longer be explained by the natural statistical uncertainty of the human experiment. According to ITU-T Recommendation P.1401 \cite{P.1401}, an objective prediction for a test item is considered an outlier if the absolute error between objective and subjective quality scores exceeds a threshold determined by a multiple of the standard error of the subjective scores. Specifically, a threshold $\tau$ is defined as
\begin{equation}
\tau =  z \cdot \sigma_i,
\label{eq:outlier}
\end{equation}
where $\sigma_i$ is the standard deviation of subjective ratings, and the Z value $z$ is typically set to 1.96 for 95\% confidence. The OR is calculated as the proportion of such outlier predictions across the dataset,
\begin{equation}
\mathrm{OR} = \frac{1}{N} \sum_{i=1}^{N} \delta_i \in [0,1].
\end{equation}
Here,
\begin{equation}
\delta_i = \quad
\begin{cases}
1, & \text{if } \left| S_{trans, i} - S_{subj, i} \right| > \tau \\
0, & \text{otherwise}
\end{cases}
\label{eq:or}
\end{equation}
where $S_{trans, i}$ is the transformed objective quality score, $S_{subj, i}$ is the subjective quality score for the \mbox{$i$-th} data point. A lower OR indicates greater agreement between the objective metric and human perception, while a higher OR reflects frequent prediction errors beyond the acceptable confidence threshold.
\end{itemize}

\section{Proposed Z-RMSE Evaluation Criterion}\label{sec:proposed_eval}
Traditional error-based evaluation metrics such as MSE and RMSE assume homoscedasticity, treating all data points as having equal variance. However, in subjective quality assessment datasets, each data point typically follows a distribution that reflects variability in human opinion. Crucially, this uncertainty is not uniform across the dataset; for example, in subjective dataset used in this work, high-quality content has lower $\sigma$, while lower-quality content has typically higher $\sigma$. The Z-score Root Mean Square Error (Z-RMSE) is a normalized error metric that accounts for variability in the subjective scores by incorporating the standard deviation at each data point. Specifically, it computes the root mean squared error of the residuals, with each residual scaled first by the corresponding subjective standard deviation. This allows Z-RMSE to account for local uncertainty in the data: prediction errors on highly consistent (low-variance) samples are penalized more heavily than those on uncertain (high-variance) samples. A lower Z-RMSE indicates that the model's predictions are closely aligned with the subjective means, especially in regions of low perceptual uncertainty.
\begin{equation}
   \text{Z-RMSE} =   \sqrt{ \frac{1}{n} \sum_{i=1}^{n} \left( \frac{S_{trans, i} - S_{subj, i}}{\sigma_i} \right)^2 } .
\end{equation}
In this formula, $S_{trans, i}$ denotes the transformed objective scores, i.e., the output of a logistic mapping that aligns the raw metric values with the scales of the subjective means. This formulation of Z-RMSE is not only intuitive but also tightly connected to a probabilistic interpretation grounded in likelihood.

To formalize this connection, consider the ground-truth perceived quality for each image as a Gaussian random variable reflecting observer uncertainty: 
\begin{equation}
    X_i \sim \mathcal{N}(\mu_i, \sigma_i)
\end{equation}
Importantly, this uncertainty differs across samples. Given this model, the likelihood of predicted mean scores $\hat{\mu}$ is
\begin{equation}
    L(\hat{\mu}) = \prod_{i=1}^{N} \frac{1}{\sqrt{2\pi\sigma_i^2}} \exp\left( -\frac{(\hat{\mu}_i - \mu_i)^2}{2\sigma_i^2} \right).
\end{equation}

For comparison, the null likelihood $L_0$ is defined as the likelihood under a perfect predictor:  $\hat{\mu}_i = \mu_i$ for all $i$. This yields
\begin{equation}
L_0 = L(\mu) = \prod_{i=1}^{N} \frac{1}.{\sqrt{2\pi\sigma_i^2}}
\end{equation}
Taking the log-likelihood ratio between the null model and the prediction model gives
\begin{equation}
\text{LLR} = \log \left( \frac{L_0}{L(\hat{\mu})} \right) = \frac{1}{2} \sum_{i=1}^{N} \left( \frac{\hat{\mu}_i - \mu_i}{\sigma_i} \right)^2
\end{equation}
This expression is proportional to the squared \mbox{Z-RMSE}
\begin{equation}
\text{LLR} = \frac{N}{2} \cdot (\text{Z-RMSE})^2.
\end{equation}
Thus, minimizing Z-RMSE is equivalent to maximizing the log-likelihood of the predictions under a Gaussian model with per-sample variances. Metrics can be compared by their Z-RMSE values and thus the one achieving the lowest value is considered to provide the most plausible generative explanation for the subjective scores, offering a model-driven measure of fit that complements rank-based indicators such as SROCC and Outlier Ratio (OR).

\section{Statistical Tests}\label{sec:stat-test}
Two statistical tests are introduced to evaluate quality metrics, providing a fundamental method to determine whether one metric outperforms another.

\subsection{Meng-Rosenthal-Rubin test}\label{sec:meng}
The Meng–Rosenthal–Rubin (MRR) test \cite{meng1992comparing} is used to assess whether the difference between the SROCC (described in Section~\ref{sec:conv_eval})  of two quality metrics is statistically significant. Note that MRR can be used for any correlation coefficient, but it was applied to SROCC due to its importance in quality assessment. Although MRR is not frequently used, it is especially useful when the correlations being compared are dependent, that is, when they share a common variable (in this case the subjective scores). Simply comparing two correlation values (e.g. 0.85 vs. 0.80) is insufficient in such cases, as the observed difference may not be statistically meaningful. The MRR test accounts for this dependency. 
Therefore, the MRR test can determine whether the SROCC correlation between the scores $S^{A}_{\text{obj}}$ of metric A and the subjective scores $S_{\text{sub}}$, is significantly different from the correlation between the scores $S^{B}_{\text{obj}}$ of metric B and the same subjective scores. The null hypothesis of the MRR test states that the SROCC between $S^{A}_{\text{obj}}$ and $S_{\text{sub}}$ is equal to the SROCC between $S^{B}_{\text{obj}}$ and $S_{\text{sub}}$, i.e., $r_{1s} = r_{2s}$. The alternative hypothesis is that these correlations differ ($r_{1s} \neq r_{2s}$ for a two-sided test, or $r_{1s} > r_{2s}$ for a one-sided test). The test statistic, $Z$, is computed as follows:
\begin{equation}
\begin{aligned}
Z &= (z_1 - z_2) \left( \frac{2(1 - r_{12}) \cdot h}{n - 3} \right)^{-0.5}, \\
z_1 &= \tanh^{-1}(r_{1s}), \\
z_2 &= \tanh^{-1}(r_{2s})
\end{aligned}
\end{equation}
Here, \( r_{1s} \), \( r_{2s} \), and \( r_{12} \) denote the SROCC between \([S_{\text{obj}}^A, S_{\text{subj}}] \), \( [S_{\text{obj}}^B, S_{\text{subj}}] \), and \( [S_{\text{obj}}^A, S_{\text{obj}}^B] \), respectively. The number of samples is given by \( n \), and the term \( h \) is a correction factor defined as
\begin{equation*}
h = \frac{1 - f \cdot \bar{r}^2}{1 - \bar{r}^2}, \quad
\bar{r}^2 = \frac{r_{1s}^2 + r_{2s}^2}{2}, \quad
f = \frac{1 - r_{12}}{2(1 - \bar{r}^2)}. 
\end{equation*}
Once the Z value is computed, the two-tailed p-value can be obtained by
\begin{equation}
    p\text{-value} = 2 \cdot \left(1 - \Phi\left(|Z|\right)\right)
\end{equation}
where $\Phi$ is the cumulative distribution function of the standard normal distribution.
The decision rule for the MRR test can be summarized as
\begin{equation}
\text{Decision} =
\begin{cases}
  1,   & \text{if } p\text{-value} < \alpha \text{ and } Z > 0 \\
 -1,   & \text{if } p\text{-value} < \alpha \text{ and } Z < 0 \\
 \text{0}, & \text{if } p\text{-value} \geq \alpha
\end{cases}
\label{eq:mrr_decision}
\end{equation}

where $\alpha$ is the significance level. A decision value of $1$ indicates that metric A correlates significantly better than metric B; $-1$ indicates the opposite; and $0$ indicates that the difference is not statistically significant.

\subsection{Wilcoxon Signed-Rank Test}\label{sec:wilcoxon}
The Wilcoxon Signed-Rank Test \cite{wilcoxon} is used to assess whether the performance differences between image quality metrics were statistically significant, determining if the observed variations reflect consistent trends rather than random fluctuations. Since each reference image is processed by several coding solutions and evaluated with corresponding quality metrics, the measurements naturally formed paired samples, i.e. each pair consists of quality scores from two different metrics for the same references. This pairing allows a direct comparison of how the metrics perform on the same set of images, thereby controlling for content-specific variations. To account for the potential non-normal distribution of the quality scores and to leverage the paired nature of the data, a Wilcoxon signed-rank test should be employed. This non-parametric test assesses whether the median of the differences between paired observations significantly differs from zero, providing a robust statistical analysis without assuming a Gaussian model. By applying the Wilcoxon signed-rank test, it is possible to evaluate if the differences between metrics' performance are not only observed empirically but also statistically significant at a given confidence level. This procedure involves the following steps:

\begin{itemize}
    \item Logistic transformation: Logistic transformation is applied to the metric values, as described in Section~\ref{sec:transformation}, to normalize the objective scores to the same range as the subjective scores. This is an essential step since a residual is calculated next.

    \item Residual calculation: After applying the logistic transformation, the residual is computed as the absolute difference between the subjective scores ($S_{\text{subj}}$) and the transformed objective scores ($S_{\text{trans}}^A$):

    \begin{equation}
        R^A = |S_{trans}^A - S_{subj}|
    \end{equation} 

    \item A paired sample Wilcoxon signed-rank test is applied. The null hypothesis of this test states that the median of the paired residual differences $d$ is zero, where $d = R^{A} - R^{B}$, with $R^{A}$ and $R^{B}$ denoting vector of the residual for metrics $A$ and $B$, respectively. Under this hypothesis, there is no significant difference between the distributions of the residuals $R^A$ and $R^B$ for the two metrics, i.e. no systematic difference in their prediction errors. Rejection of the null hypothesis indicates that the median of the paired differences $d_i$ significantly deviates from zero, implying that one metric consistently yields smaller errors than the other at the chosen confidence level. 

    \item Effect size: The test statistic $W$ is computed as the sum of the signed ranks of the absolute differences between paired observations, excluding zero differences. $W$ is then converted into a standardized $Z$-value using a normal approximation, which is appropriate when the number of non-zero pairs, $n$, is sufficiently large. The $Z$-value is calculated as:
    \begin{equation}
    \begin{aligned}
    Z &= \frac{W - \mu_w}{\sigma_w}, \\
    \mu_w &= \frac{n(n+1)}{2}, \\
    \sigma_w &= \sqrt{\frac{n(n+1)(2n+1)}{24}}
    \end{aligned}
    \end{equation}
where $ \mu_w $ and $\sigma_w $ are the mean and standard deviation of the Wilcoxon distribution under the null hypothesis, and $n$ is the number of non-zero differences. 

This standardized $Z$-value is then used to compute the effect size, providing an interpretable measure of the strength of the observed difference between the paired samples:
\begin{equation}
    r = \frac{Z}{\sqrt{N}}
\end{equation}

    \item p-value evaluation: If the \textit{p}-value from the Wilcoxon signed-rank test, is less than the effect size threshold $r$, this indicates a statistically significant difference between the two metrics. In this case, their median prediction residuals, $ \hat{R}^{\text{A}} $ and $ \hat{R}^{\text{B}}$, are compared. If $\hat{R}^{\text{A}} < \hat{R}^{\text{B}}$, a value of $1$ is assigned, indicating that metric $A$ performs better. Otherwise, a value of $-1$ is assigned, indicating that metric $A$ performs worse. If the \textit{p}-value is greater than or equal to the threshold $r$, no statistically significant difference is observed, and a value of $0$ is assigned, denoting that the two metrics are statistically indistinguishable.
        
This decision rule can be summarized as:
\begin{equation}
\begin{aligned}
   \text{Decision} = 
   \begin{cases}
     1, & \text{if } p\text{-value} < r \text{ and } \hat{R}^{\text{A}} < \hat{R}^{\text{B}} \\
    -1, & \text{if } p\text{-value} < r \text{ and } \hat{R}^{\text{A}} \geq \hat{R}^{\text{B}} \\
    \text{0}, & \text{if } p\text{-value} \geq r 
  \end{cases}
\end{aligned}
\label{eq:wc_decision}
\end{equation}
  
\end{itemize}

\begin{table*}[!htbp]
\caption{Performance evaluation of quality metrics on the full-resolution image dataset sorted from the lowest to the highest SROCC (\textbf{All}). Experimental results are reported for the entire dataset (denoted as \textbf{All}) as well as for two subsets \textbf{HF} and \textbf{MF} with perceived distortion in the range up to 1 JND and in the range above 1 JND, respectively. The best four quality metrics are highlighted in bold for each range.}
\begin{adjustbox}{width=\textwidth, center}
\centering
\small
\renewcommand{\arraystretch}{1}
\begin{tabular}{l|ccc|ccc|ccc|ccc}
\hline
\specialrule{0.2em}{\abovetopsep}{\belowbottomsep}
\multirow{2}{*}{\textbf{Quality Metric}} 
& \multicolumn{3}{c|}{\textbf{PLCC$\uparrow$}} 
& \multicolumn{3}{c|}{\textbf{SROCC$\uparrow$}} 
& \multicolumn{3}{c|}{\textbf{RMSE$\downarrow$}} 
& \multicolumn{3}{c}{\textbf{KT$\uparrow$}}
\\
\cline{2-13}
& \multicolumn{1}{c}{\textbf{All}} & \multicolumn{1}{c}{\textbf{HF}} & \multicolumn{1}{c|}{\textbf{MF}}  
& \multicolumn{1}{c}{\textbf{All}} & \multicolumn{1}{c}{\textbf{HF}} & \multicolumn{1}{c|}{\textbf{MF}} 
& \multicolumn{1}{c}{\textbf{All}} & \multicolumn{1}{c}{\textbf{HF}} & \multicolumn{1}{c|}{\textbf{MF}}
& \multicolumn{1}{c}{\textbf{All}} & \multicolumn{1}{c}{\textbf{HF}} & \multicolumn{1}{c}{\textbf{MF}} \\

\hline
\specialrule{0.2em}{\abovetopsep}{\belowbottomsep}

\textbf{UQI}
&0.319 &0.004 &0.412   &0.518 &0.378 &0.456  &0.885 &0.994 &0.810   &0.400 &0.278 &0.326             \\ 

\textbf{FLIP}
&0.672 &0.433 &0.378   &0.631 &0.524 &0.154  &0.692 &0.730 &0.667   &0.456 &0.389 &0.140             \\

\textbf{TOPIQ}
&0.779 &0.106 &0.667   &0.763 &0.017 &0.662 &0.585 &0.596 &0.578 &0.556 &0.002 &0.475             \\

\textbf{DISTS}
&0.816 &0.437 &0.722    &0.801 &0.432 &0.694 &0.539 &0.499 &0.563 &0.604 &0.301 &0.505             \\ 

\textbf{PSNRY}   
&0.804 &0.734 &0.387    &0.812 &0.784 &0.400 &0.554 &0.409 &0.628 &0.626 &0.608 &0.285             \\ 

\textbf{A-DISTS}
&0.857 &0.551 &0.753    &0.848 &0.520 &0.743 &0.481 &0.423 &0.513 &0.654 &0.368 &0.546             \\ 

\textbf{VSI}
&0.845 &0.514 &0.728    &0.848 &0.595 &0.703 &0.499 &0.434 &0.535 &0.661 &0.433 &0.516             \\ 

\textbf{FSIM}
&0.855 &0.554 &0.712    &0.861 &0.604 &0.712 &0.483 &0.424 &0.517 &0.680 &0.449 &0.535             \\

\textbf{LPIPS}
&0.856 &0.743 &0.600    &0.867 &0.814 &0.612 &0.482 &0.388 &0.532 &0.689 &0.620 &0.439             \\

\textbf{FSIMc}
&0.866 &0.568 &0.726  &0.875 &0.633 &0.734                    &0.465 &0.399 &0.502                    &0.698 &0.470 &0.557             \\

\textbf{Haar-PSI}
&0.863 &0.784 &0.565   &0.875 &0.822 &0.583                     &0.471 &0.300 &0.551                    &0.700 &0.667 &0.412             \\

\textbf{PSNR-HVS}   
&0.867  &0.746 &0.606   &0.877  &0.774 &0.621                   &0.465  &0.308 &0.540           &0.701  &0.616 &0.446  \\

\textbf{VMAF}
&0.883 &0.649 &0.722     &0.889 &0.635 &0.729                    &0.437 &0.355 &0.482                     &0.703 &0.454 &0.538             \\ 

\textbf{GMSD}
&0.882 &0.755 &0.658    &0.892 &0.771 &0.676                    &0.439 &0.296 &0.509                    &0.717 &0.607 &0.487            \\

\textbf{BUTTERAUGLI}
&0.881 &0.776 &0.616    &0.893 &0.857 &0.640                    &0.442 &0.280 &0.517                    &0.719 &0.674 &0.454             \\ 

\textbf{SSIMULACRA2}
&0.893 &0.806 &0.657     &0.905 &0.831 &0.687                    &0.420 &0.242 &0.500                    &0.737 &0.679 &0.490             \\

\textbf{VIF}
&0.894 &0.830 &0.662     &0.905 &0.839 &0.684                    &0.418 &0.238 &0.499                    &0.740 &0.694 &0.492             \\

\textbf{SSIMULACRA1}
&0.897  &0.806 &0.681     &0.907  &0.854 &0.691                          &0.412  &0.262 &0.482            &0.741  &0.682 &0.505  \\

\textbf{HDR-VDP-2} 
&0.900  &0.725 &0.733     &0.908  &0.807 &0.742                          &0.408  &0.327 &0.451            &0.747  &0.629 &0.562  \\

\textbf{PieAPP}
&0.906 &0.692 &\textbf{0.777}      &0.909 &0.740 &\textbf{0.764}         &0.395 &0.325 &\textbf{0.432}            &0.743 &0.581 &\textbf{0.571}             \\ 

\textbf{SSIM}
&0.900 &0.803 &0.690      &0.911 &\textbf{0.864} &0.704                    &0.407 &0.265 &0.474                    &0.746 &0.695 &0.512             \\

\textbf{VMAF-NEG}
&0.906 &0.775 &\textbf{0.741 }     &0.915 &0.810 &0.743                    &0.395 &0.296 &0.446                    &0.744 &0.604 &0.550            \\ 

\textbf{NLPD}
&0.904 &\textbf{0.848} &0.685      &0.917 &\textbf{0.873} &0.714             &0.399 &\textbf{0.214} &0.479       &0.757 &\textbf{0.729} &0.515  \\

\textbf{MS-SSIM}
&\textbf{0.915} &0.831 &0.735  &\textbf{0.927} &0.855 &0.763  &\textbf{0.376} &0.220 &0.446  &\textbf{0.770} &\textbf{0.695} &0.566  \\

\textbf{HDR-VDP-3}
&\textbf{0.917} &\textbf{0.833} &0.738  &\textbf{0.929}  &\textbf{0.873} &\textbf{0.768}  &\textbf{0.372}  &\textbf{0.219} &\textbf{0.441}  &\textbf{0.778}  &\textbf{0.714 }&\textbf{0.579}  \\

\textbf{IW-SSIM}
&\textbf{0.940} &\textbf{0.851} &\textbf{0.826} &\textbf{0.944} &\textbf{0.867} &\textbf{0.825} &\textbf{0.317} &\textbf{0.196} &\textbf{0.373} &\textbf{0.802} &\textbf{0.715} &\textbf{0.635} \\ 

\textbf{CVVDP}
&\textbf{0.958} &\textbf{0.842} &\textbf{0.888}  &\textbf{0.960} &0.852 &\textbf{0.893}  &\textbf{0.265} &\textbf{0.184} &\textbf{0.305}  &\textbf{0.838} &\textbf{0.715 }&\textbf{0.722} \\

\hline
\specialrule{0.2em}{\abovetopsep}{\belowbottomsep}
\end{tabular}
\label{tbl_performance1}
\end{adjustbox}
\end{table*}

\section{Experimental Results}\label{sec:exp-results}
This section evaluates the metrics presented in Section~\ref{obj_quality_metrics} using the proposed evaluation criteria and assesses the statistical significance of the results. A subjective dataset that is introduced in Section~\ref{sec:aic3_dataset}. The performance of the metrics is then analyzed and statistically assessed in Sections~\ref{lbl-performance} and~\ref{lbl-statisticalTest}, respectively. The impact of cropping is examined in Section~\ref{lbl-resolution}. The subjective methodology and the processing to obtain the perceived impairments in JND units is described in Section~\ref{sec:aic3_processing}.

\subsection{Overall Evaluation} \label{lbl-performance}
This section evaluates the performance of the metrics using the full-resolution images and the subjective scores available in the JPEG AIC-3 dataset. Note that only crops were shown to the subjects during subjective viewing and thus, it is assumed that the crop image quality is similar to the entire image quality. The impact of this choice is further studied in Section~\ref{lbl-resolution}. The experimental results are reported for three cases:
\begin{itemize}
\item \textbf{All}: all stimuli contained in the JPEG AIC-3 dataset. This corresponds to the entire set of 300 scores, covering the range from 0 to 3.8 JND units.
\item High Fidelity (\textbf{HF}): subset of stimuli characterized by perceived impairments within the interval $[0, 1]$ JND units, comprising 115 scores. This interval corresponds to conditions in which compression artifacts remain at or below the visibility threshold for an average human observer, a regime for which, to date, no studies have systematically assessed the performance of the quality metrics. This region is critical for professional and quality-sensitive applications, such as digital archiving or medical imaging, where any perceivable degradation is unacceptable and even sub-threshold distortions can have practical consequences.
\item Medium Fidelity (\textbf{MF}): subset of stimuli defined by the perceived impairments in the range $[1, +\infty]$ in JND units, corresponding to 185 scores. In this interval, distortions are perceptible to the average human observer, and thus constitute the range in which existing quality metrics may have been validated with scores obtained from other methodologies. This region is highly relevant for applications where perceptual quality is important but some (small) degree of visible degradation is acceptable, such as web delivery.
\end{itemize}
This separation at 1 JND, the perceptual visibility threshold, enables targeted analysis of metric performance in subtle versus pronounced distortion regimes, offering clearer insight into their suitability for different use cases.

An additional consideration is that the transformation described in Section~\ref{sec:transformation}, is applied globally. Specifically, for each quality metric, objective scores are computed using the entire set of stimuli and sources, and the transformation parameters are estimated from all available subjective scores (\textbf{All} case), independently of the evaluation measures and fidelity ranges, thereby ensuring fairness and consistency in the comparison.
\subsubsection{PLCC, SROCC, RMSE and KT}
The experimental results, assessed using PLCC, SROCC, RMSE, and KT (described in Section~\ref{sec:conv_eval}), are summarized in Table~\ref{tbl_performance1}. Based on these results, the following key observations can be made:
\begin{itemize}
\item The performance metrics PLCC, SROCC, RMSE, and KT exhibit strong mutual correlation. According to these criteria, the top-performing quality metrics for the \textbf{All} case are CVVDP, IW-SSIM, MS-SSIM and HDR-VDP-2/3.
\item Learning-based quality metrics generally perform worse than conventional metrics. This is likely because most of them do not include high-to-visually-lossless quality ranges in their training data and are trained using scores obtained from Absolute Category Ratings, rather than pairwise comparisons (with the exception of PieAPP).
\item Metrics that perform best for the entire range tend also to rank among the best for the \textbf{HF} and/or \textbf{MF} subsets. Notably, NLPD is among the best in the \textbf{HF} range, while PieAPP is among the best in the \textbf{MF} range.
\item For the \textbf{HF} subset, PLCC, SROCC, and KT values are generally lower than for the \textbf{MF} subset, as expected given the greater difficulty of the \textbf{HF} range. Conversely, RMSE is lower for \textbf{HF} than for \textbf{MF}, reflecting the smaller magnitude of errors in the high-fidelity regime.
\end{itemize}
\subsubsection{OR, PWRC, and Z-RMSE}
\begin{figure*}[!htbp]
\centering
\begin{subfigure}{0.32\textwidth}
    \includegraphics[width=\linewidth]{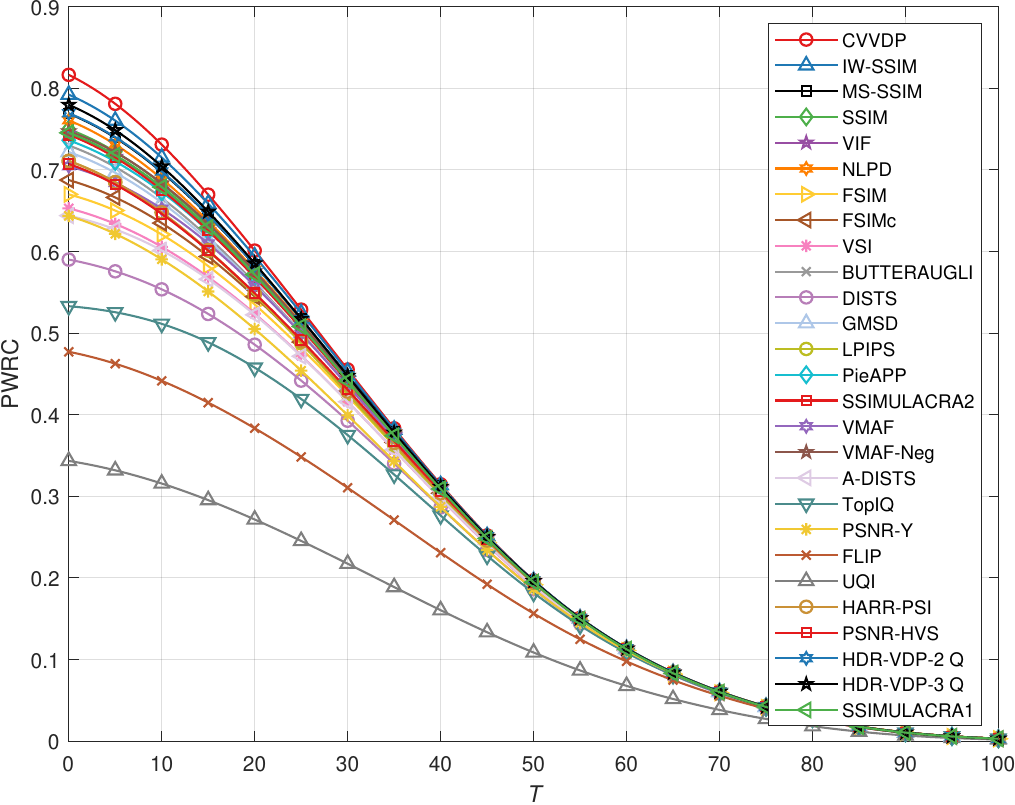}
    \caption{Overall dataset (\textbf{All})}
    \label{fig:pwrc_Ovr}
\end{subfigure}
\hfill
\begin{subfigure}{0.32\textwidth}
    \includegraphics[width=\linewidth]{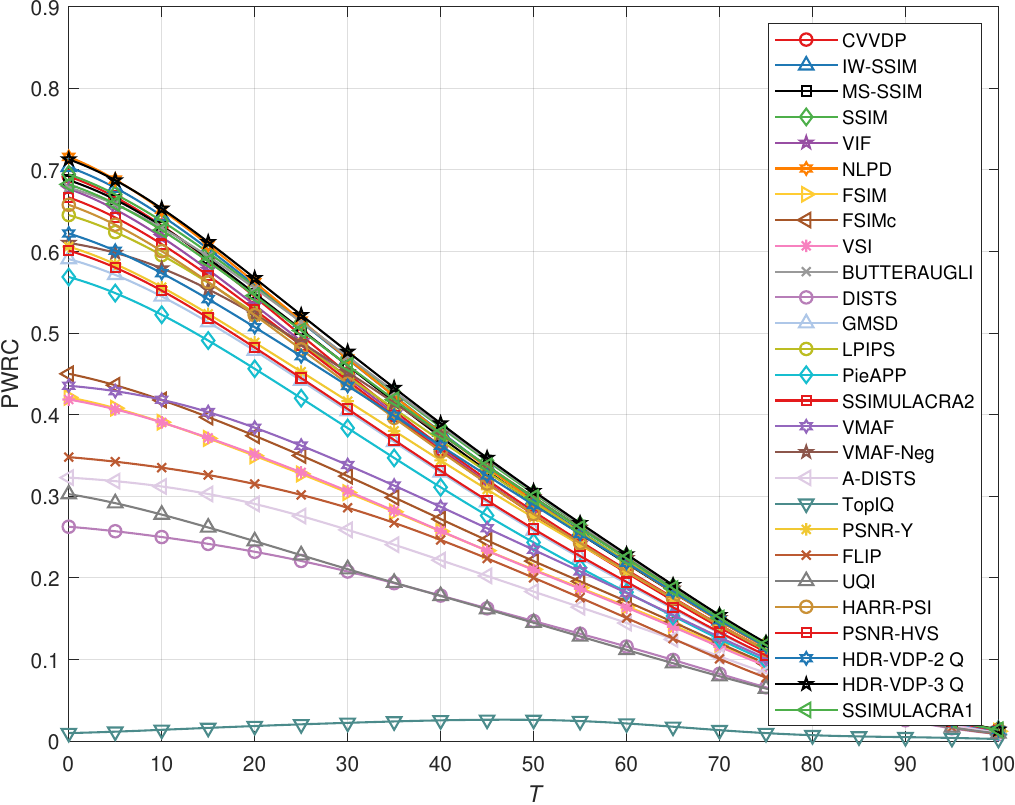}
    \caption{JND in $ HF = [0, 1]$ quality range}
    \label{fig:pwrc_f1}
\end{subfigure}
\hfill
\begin{subfigure}{0.32\textwidth}
    \includegraphics[width=\linewidth]{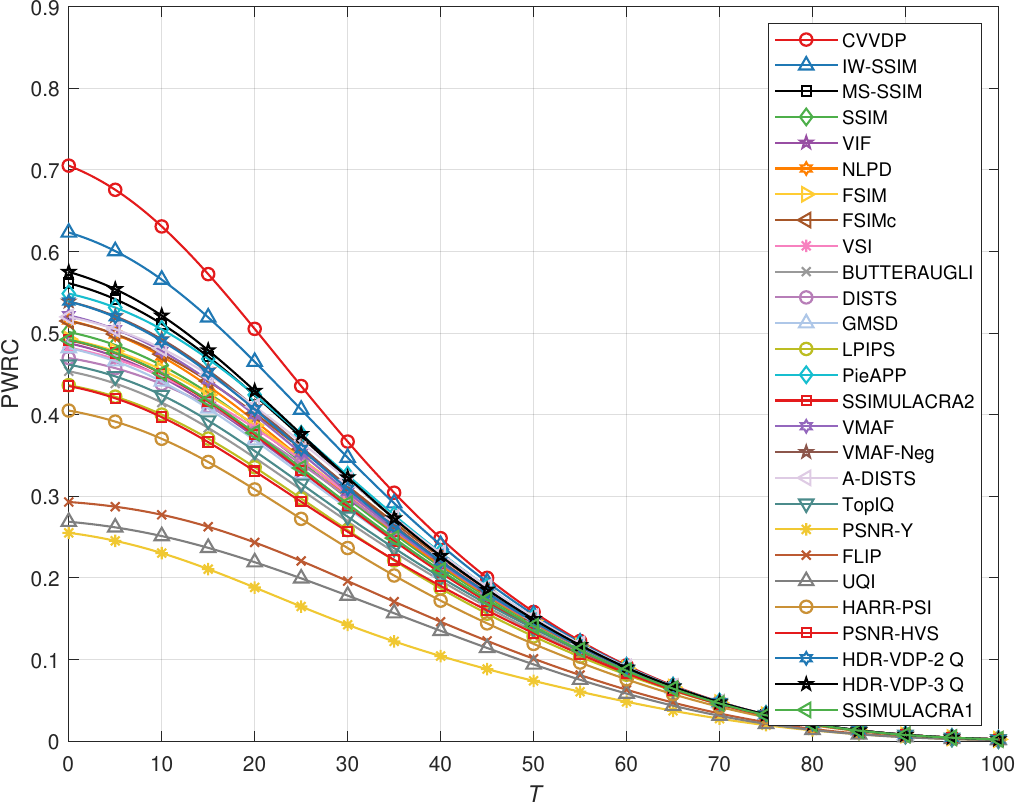}
    \caption{JND in $MF(1, +\infty]$ quality range}
    \label{fig:pwrc_f2}
\end{subfigure}
\caption{Performance of the PWRC evaluation criteria for different JND quality ranges}
\label{Fig_pwrc}
\end{figure*}

The experimental results, assessed using OR, PWRC and Z-RMSE (described in Section~\ref{sec:conv_eval} and Section~\ref{sec:proposed_eval}), are summarized in Table~\ref{tbl_performance2}. All of these evaluation measures use confidence intervals or the standard deviation directly. Additionally, the graphical SA-ST curve results of the PWRC evaluation criteria are presented in Figure \ref{Fig_pwrc}. This figure represents the performance of the quality metrics across different sensory threshold (ST). A higher PWRC value suggests that the metric aligns more closely with human visual perception. However, as ST continues to increase, the SA for all quality metrics approaches zero, indicating that the differences in metric performance become indistinguishable. Therefore, the area under the curve (AUC) for each metric is reported in Table \ref{tbl_performance1}. Based on these results, the following key observations can be made:
\begin{itemize}
\item For the OR and Z-RMSE performance metrics, CVVDP is always the best but the third or second best can be different. Interestingly, PSNRY despite its simplicity still has a good performance for Z-RMSE, due to its reliability (since it does not require models of perception, or training data, which may fail for some cases).  
\item OR, PWRC and Z-RMSE values are consistently higher in the \textbf{HF} subset compared to the \textbf{MF} subset. Z-RMSE doesn't follow the same behavior as RMSE since the standard deviation is rather small for impairment values below 1 JND. For PWRC, as shown in Figure~\ref{Fig_pwrc}, \textbf{MF} and \textbf{HF} values are lower than the \textbf{All} set at zero threshold.
\item Another interesting point is the high values for the OR and Z-RMSE evaluation measures. This comes from the fact that the standard deviation using this methodology is rather small and thus the quality metric often deviates from the target quality score $\pm \sigma$. This means that there is still a large room for improvement in terms of the quality metrics accuracy.
\end{itemize}

\begin{table*}[!htpb]
\centering
\small
\caption{Performance evaluation of quality metrics on the JPEG AIC full-resolution image dataset sorted as in previous Table~\ref{tbl_performance1} with SROCC. Experimental results are reported for the entire dataset (denoted as \textbf{All}) as well as for two subsets \textbf{HF} and \textbf{MF} with perceived distortion in the range up to 1 JND and in the range above 1 JND, respectively. The best four quality metrics are highlighted for each case.}

\renewcommand{\arraystretch}{1}
\begin{tabular}{l|ccc|ccc|ccc}
\hline
\specialrule{0.2em}{\abovetopsep}{\belowbottomsep}
\multirow{2}{*}{\textbf{Quality Metric}} 
& \multicolumn{3}{c|}{\textbf{OR$\downarrow$}}
& \multicolumn{3}{c|}{\textbf{PWRC$\uparrow$}}
& \multicolumn{3}{c}{\textbf{Z-RMSE$\downarrow$}} \\
\cline{2-10}

& \multicolumn{1}{c}{\textbf{All}} & \multicolumn{1}{c}{\textbf{HF}} & \multicolumn{1}{c|}{\textbf{MF}}  
& \multicolumn{1}{c}{\textbf{All}} & \multicolumn{1}{c}{\textbf{HF}} & \multicolumn{1}{c|}{\textbf{MF}}  
& \multicolumn{1}{c}{\textbf{All}} & \multicolumn{1}{c}{\textbf{HF}} & \multicolumn{1}{c}{\textbf{MF}}

\\
\hline
\specialrule{0.2em}{\abovetopsep}{\belowbottomsep}

\textbf{UQI}           &0.913  &0.991 &0.864            &2.51  &4.52 &1.80            &422.91  &682.98 &8.44         \\ 

\textbf{FLIP}          &0.856  &0.895 &0.832           &3.49  &5.40 &1.98      &100.22  &161.59 &7.51    \\

\textbf{TOPIQ}         &0.790  &0.869 &0.740            &3.95  &0.20 &3.06     &605.93  &978.62 &7.11     \\

\textbf{DISTS}         &0.846  &0.886 &0.821            &4.34  &4.04 &3.14          &252.53  &407.78 &7.05             \\ 

\textbf{PSNRY}         &0.843  &0.895 &0.810            &4.70  &9.05 &1.67        &\textbf{13.36}  &\textbf{19.39} &7.46       \\ 

\textbf{A-DISTS}       &0.850  &0.895 &0.821         &4.73  &5.02 &3.46            &221.26  &357.27 &6.55    \\ 

\textbf{VSI}           &0.883  &0.913 &0.864            &4.79  &6.33 &3.21         &122.26  &197.28 &6.75          \\ 

\textbf{FSIM}          &0.796  &0.843 &0.767           &4.91  &6.34 &3.28        &71.32  &114.91 &6.41          \\

\textbf{LPIPS}         &0.833  &0.860 &0.816          &5.18  &9.67 &2.89          &76.88  &123.93 &6.15           \\

\textbf{FSIMc}         &0.820  &0.826 &0.816         &5.03  &6.78 &3.42          &70.89  &114.22 &6.34       \\

\textbf{Haar-PSI}      &0.793  &0.826 &0.772         &5.18  &9.78 &2.68     &56.68  &91.17 &6.61    \\

\textbf{PSNR-HVS}      &0.826  &0.869 &0.800         &5.16  &8.98 &2.87         &50.63  &81.37 &6.45      \\ 

\textbf{VMAF}          &0.800  &0.895 &0.740       &5.16  &6.73 &3.46              &134.69  &217.43 &5.75  \\ 

\textbf{GMSD}          &0.773  &0.800 &0.756        &5.27  &8.86 &3.19               &16.87  &\textbf{26.16} &6.03        \\

\textbf{BUTTERAUGLI} &0.843 &0.808 &0.864           &5.32  &10.21 &3.00         &25.94  &41.12 &6.33    \\ 

\textbf{SSIMULACRA2}   &0.740  &\textbf{0.756} &0.729         &5.41  &9.92 &3.26      &47.63  &76.54 &6.08    \\

\textbf{VIF}           &0.776  &0.808 &0.756         &5.42  &10.06 &3.23             &40.34  &64.72 &6.01          \\

\textbf{SSIMULACRA1}   &0.770  &0.773 &0.767         &5.43  &10.19 &3.26      &64.51  &103.90 &6.09     \\

\textbf{HDR-VDP-2}   &\textbf{0.683}  &0.686 &\textbf{0.681}        &5.41  &9.32 &3.55      &35.04  &56.17 &5.51          \\

\textbf{PieAPP}        &\textbf{0.743}  &0.826 &\textbf{0.691}        &5.37  &8.50 &3.64       &54.86  &88.35 &\textbf{5.33}   \\ 

\textbf{SSIM}          &0.790  &0.773 &0.800        &5.46  &\textbf{10.37} &3.33             &60.38  &97.23 &5.92        \\

\textbf{VMAF-NEG}      &0.753  &0.817 &0.713        &5.46  &9.35 &3.56                   &\textbf{22.44}  &35.60 &5.34  \\ 

\textbf{NLPD}          &0.753  &0.782 &0.735        &5.53  &\textbf{10.62} &3.41            &53.33  &85.82 &5.77            \\

\textbf{MS-SSIM}       &0.770  &0.800 &0.751        &\textbf{5.60}  &10.26 &\textbf{3.70}       & 27.79  &44.38 &5.35          \\

\textbf{HDR-VDP-3}   &0.783  &0\textbf{.760} &0.794         &\textbf{5.67}  &\textbf{10.62} &\textbf{3.78}     &\textbf{12.30}  &\textbf{18.70} &\textbf{5.30}  \\

\textbf{IW-SSIM}       &\textbf{0.726}  &\textbf{0.765} &\textbf{0.702}  &\textbf{5.76} &\textbf{10.48} &\textbf{4.10}  &31.51  &50.58 &\textbf{4.50}    \\ 

\textbf{CVVDP}         &\textbf{0.593}  &\textbf{0.695} &\textbf{0.529}  &\textbf{5.92} &10.29 &\textbf{4.60}    &\textbf{9.45}  &\textbf{14.50} &\textbf{3.77}\\

\hline
\specialrule{0.2em}{\abovetopsep}{\belowbottomsep}
\end{tabular}
\label{tbl_performance2}
\end{table*}

\subsubsection{RMSE vs Z-RMSE}
\begin{figure}
    \centering
    \includegraphics[width=1\linewidth]{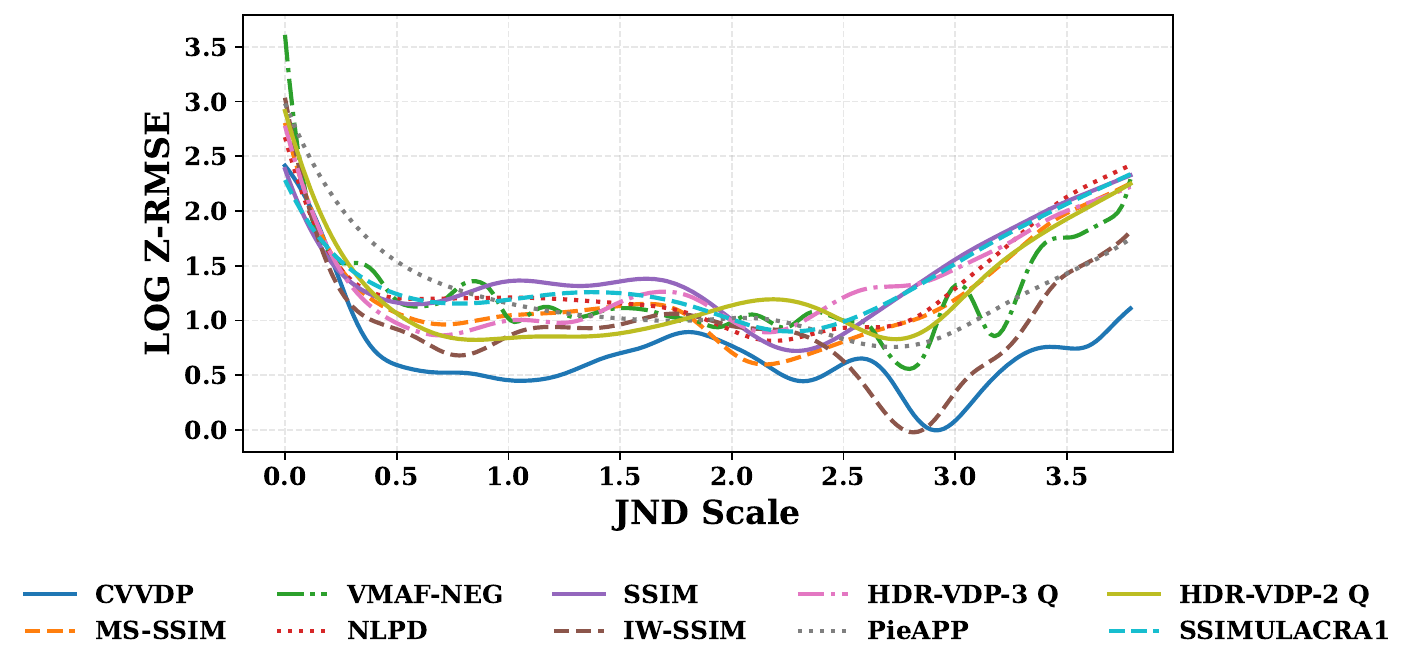}
    \caption{Z-RMSE evolution (on vertical log scale) from low to higher average scores.}
    \label{fig:zrmse_eval}
\end{figure}

A more detailed comparison between RMSE and Z-RMSE is conducted by analyzing the evolution of each metric as a function of the perceptual scores. To avoid the limitations of coarse binning and linear interpolation, the relationship is modeled using a non-parametric Nadaraya–Watson kernel regression with a Gaussian kernel and automatic bandwidth selection (least-squares cross-validation). This approach provides a smooth estimate of the conditional expectation of each metric RMSE across the entire range of JND values, without imposing strong parametric assumptions on the data distribution. The experimental results, presented in Figure~\ref{fig:zrmse_eval}, show the regression curves, where RMSE increases monotonically with perceptual distortion, reflecting the higher magnitude of errors at larger distortion levels. In contrast, Z-RMSE exhibits the opposite trend, attributable to the substantially lower standard deviations observed for low JNDs. This characteristic aligns Z-RMSE with the behavior of other evaluation measures such as PLCC, PWRC, and SROCC, highlighting its consistency with established correlation-based criteria.

\begin{figure}
    \centering
    \includegraphics[width=1\linewidth]{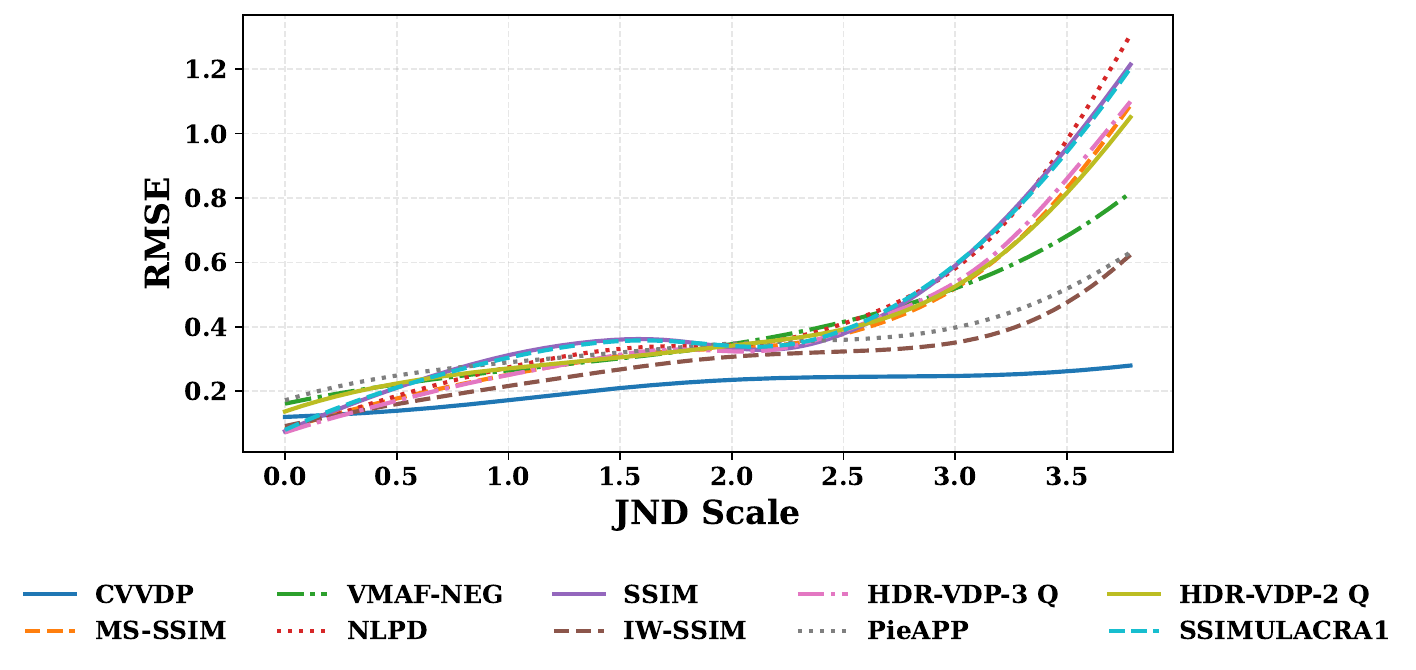}
    \caption{RMSE evolution from low to higher scores.}
    \label{fig:rmse_eval}
\end{figure}

\subsubsection{Analysis per Source and Codec}
Additionally, Figure~\ref{Fig_model_vs_pref} illustrates the relationship between subjective scores (perceived impairment in JND units) and objective scores obtained after the transformation of (\ref{equ-logistic}) described in Section \ref{sec:transformation}, for six objective quality metrics: PSNRY, MS-SSIM, IW-SSIM, CVVDP, SSIMULACRA2, and HDR-VDP-3. Each data point is accompanied by its standard deviation. As shown, the standard deviation, or uncertainty, tends to be lower for samples with smaller perceived impairment in JND units and higher for those with larger JND units. As can be observed, there is a significant deviation from the identity line (ideal case) for some sources, e.g. SRC02, when for other sources most of the quality metrics can do a good prediction (e.g. SRC10). Another interesting conclusion is that the performance of quality metrics changes significantly depending on the codecs (and thus type of degradations), especially for the JPEG AI learning-based codec where there is a clear deviation from the ideal case. 

\subsection{Statistical Test Results}\label{lbl-statisticalTest}
The results of the MRR test are summarized in Table \ref{tbl_MRR_test} with metrics sorted from the highest to lowest SROCC. Each entry reflects the result of a pairwise statistical comparison between the metric listed in the row and the metric in the column: a black cell denotes that the column metric performs significantly better, a white cell indicates that it performs significantly worse, and a gray cell corresponds to the absence of a statistically significant difference. In accordance with the decision rule defined in equation \ref{eq:mrr_decision} (see Section~\ref{sec:meng}), the CVVDP metric demonstrates statistically significant superiority over all other metrics considered. The IW-SSIM comes next since it has statistically significant better performance for all metrics except CVVDP, followed by HDR-VDP-3.

The results of the statistical pairwise Wilcoxon Signed-Rank test are summarized in Table \ref{tbl-Wincox} sorted also from the highest to the lowest SROCC. As shown in the table, the CVVDP quality metric has statistically superiority when paired to all other metrics, according to the decision rule defined in equation (\ref{eq:mrr_decision}) (see Section~\ref{sec:wilcoxon}). Additionally, IW-SSIM has statistically significant better performance for all metrics except CVVDP. This is followed by a group of 5 metrics: MS-SSIM, HDR-VDP-3, NLPD and HDR-VDP-2 for which no statistically significant difference among any pair of them. As expected, the differences between the MRR and Wilcoxon Signed-Rank Test are small and only occur for some specific pairs of image quality metrics.

\begin{table}[t]
\centering
\renewcommand{\arraystretch}{2}
\newcommand{\cellheader}[1]{\scriptsize\textbf{#1}}
\newcommand{\rot}[1]{\rotatebox{90}{\Large\textbf{#1}}}
\newcommand{\negat}{\cellcolor{black!80}}
\newcommand{\zer}{\cellcolor{gray!30}}
\newcommand{\pos}{\cellcolor{white}}
\tiny
\begin{adjustbox}{max width=\columnwidth}
\begin{tabular}{
  >{\cellheader}l *{27}{>{\centering\arraybackslash}p{0.25cm}} 
}
\toprule
& \rot{01} & \rot{02} & \rot{03} & \rot{04} & \rot{05} & \rot{06} & \rot{07} & \rot{08} & \rot{09} & \rot{10} & \rot{11} & \rot{12} & \rot{13} & \rot{14} & \rot{15} & \rot{16} & \rot{17} & \rot{18} & \rot{19} & \rot{20} & \rot{21} & \rot{22} & \rot{23} & \rot{24} & \rot{25} & \rot{26} & \rot{27} \\
\midrule

{\Large\textbf{01-CVVDP}}	&	\zer	&	\pos	&	\pos	&	\pos	&	\pos	&	\pos	&	\pos	&	\pos	&	\pos	&	\pos	&	\pos	&	\pos	&	\pos	&	\pos	&	\pos	&	\pos	&	\pos	&	\pos	&	\pos	&	\pos	&	\pos	&	\pos	&	\pos	&	\pos	&	\pos	&	\pos	&	\pos	\\
{\Large\textbf{02-IW-SSIM}}	&	\negat	&	\zer	&	\pos	&	\pos	&	\pos	&	\pos	&	\pos	&	\pos	&	\pos	&	\pos	&	\pos	&	\pos	&	\pos	&	\pos	&	\pos	&	\pos	&	\pos	&	\pos	&	\pos	&	\pos	&	\pos	&	\pos	&	\pos	&	\pos	&	\pos	&	\pos	&	\pos	\\
{\Large\textbf{03-HDR-VDP-3}}	&	\negat	&	\negat	&	\zer	&	\zer	&	\zer	&	\pos	&	\pos	&	\pos	&	\pos	&	\pos	&	\pos	&	\pos	&	\pos	&	\pos	&	\pos	&	\pos	&	\pos	&	\pos	&	\pos	&	\pos	&	\pos	&	\pos	&	\pos	&	\pos	&	\pos	&	\pos	&	\pos	\\
{\Large\textbf{04-MS-SSIM}}	&	\negat	&	\negat	&	\zer	&	\zer	&	\zer	&	\zer	&	\pos	&	\pos	&	\pos	&	\pos	&	\pos	&	\pos	&	\pos	&	\pos	&	\pos	&	\pos	&	\pos	&	\pos	&	\pos	&	\pos	&	\pos	&	\pos	&	\pos	&	\pos	&	\pos	&	\pos	&	\pos	\\
{\Large\textbf{05-NLPD}}	&	\negat	&	\negat	&	\zer	&	\zer	&	\zer	&	\zer	&	\zer	&	\zer	&	\zer	&	\zer	&	\pos	&	\pos	&	\pos	&	\pos	&	\pos	&	\pos	&	\pos	&	\pos	&	\pos	&	\pos	&	\pos	&	\pos	&	\pos	&	\pos	&	\pos	&	\pos	&	\pos	\\
{\Large\textbf{06-VMAF-NEG}}	&	\negat	&	\negat	&	\negat	&	\zer	&	\zer	&	\zer	&	\zer	&	\zer	&	\zer	&	\zer	&	\zer	&	\zer	&	\pos	&	\pos	&	\pos	&	\pos	&	\pos	&	\pos	&	\pos	&	\pos	&	\pos	&	\pos	&	\pos	&	\pos	&	\pos	&	\pos	&	\pos	\\
{\Large\textbf{07-SSIM}}	&	\negat	&	\negat	&	\negat	&	\negat	&	\zer	&	\zer	&	\zer	&	\zer	&	\zer	&	\zer	&	\zer	&	\zer	&	\zer	&	\zer	&	\pos	&	\pos	&	\pos	&	\pos	&	\pos	&	\pos	&	\pos	&	\pos	&	\pos	&	\pos	&	\pos	&	\pos	&	\pos	\\
{\Large\textbf{08-PieAPP}}	&	\negat	&	\negat	&	\negat	&	\negat	&	\zer	&	\zer	&	\zer	&	\zer	&	\zer	&	\zer	&	\zer	&	\zer	&	\zer	&	\zer	&	\zer	&	\pos	&	\pos	&	\pos	&	\pos	&	\pos	&	\pos	&	\pos	&	\pos	&	\pos	&	\pos	&	\pos	&	\pos	\\
{\Large\textbf{09-HDR-VDP-2}}	&	\negat	&	\negat	&	\negat	&	\negat	&	\zer	&	\zer	&	\zer	&	\zer	&	\zer	&	\zer	&	\zer	&	\zer	&	\zer	&	\zer	&	\zer	&	\pos	&	\pos	&	\pos	&	\pos	&	\pos	&	\pos	&	\pos	&	\pos	&	\pos	&	\pos	&	\pos	&	\pos	\\
{\Large\textbf{10-SSIMULACRA1}}	&	\negat	&	\negat	&	\negat	&	\negat	&	\zer	&	\zer	&	\zer	&	\zer	&	\zer	&	\zer	&	\zer	&	\zer	&	\zer	&	\zer	&	\pos	&	\pos	&	\pos	&	\pos	&	\pos	&	\pos	&	\pos	&	\pos	&	\pos	&	\pos	&	\pos	&	\pos	&	\pos	\\
{\Large\textbf{11-VIF}}	&	\negat	&	\negat	&	\negat	&	\negat	&	\negat	&	\zer	&	\zer	&	\zer	&	\zer	&	\zer	&	\zer	&	\zer	&	\zer	&	\zer	&	\zer	&	\pos	&	\pos	&	\pos	&	\pos	&	\pos	&	\pos	&	\pos	&	\pos	&	\pos	&	\pos	&	\pos	&	\pos	\\
{\Large\textbf{12-SSIMULACRA2}}	&	\negat	&	\negat	&	\negat	&	\negat	&	\negat	&	\zer	&	\zer	&	\zer	&	\zer	&	\zer	&	\zer	&	\zer	&	\zer	&	\zer	&	\zer	&	\pos	&	\pos	&	\pos	&	\pos	&	\pos	&	\pos	&	\pos	&	\pos	&	\pos	&	\pos	&	\pos	&	\pos	\\
{\Large\textbf{13-BUTTERAUGLI}}	&	\negat	&	\negat	&	\negat	&	\negat	&	\negat	&	\negat	&	\zer	&	\zer	&	\zer	&	\zer	&	\zer	&	\zer	&	\zer	&	\zer	&	\zer	&	\zer	&	\zer	&	\zer	&	\zer	&	\pos	&	\pos	&	\pos	&	\pos	&	\pos	&	\pos	&	\pos	&	\pos	\\
{\Large\textbf{14-GMSD}}	&	\negat	&	\negat	&	\negat	&	\negat	&	\negat	&	\negat	&	\zer	&	\zer	&	\zer	&	\zer	&	\zer	&	\zer	&	\zer	&	\zer	&	\zer	&	\pos	&	\pos	&	\zer	&	\zer	&	\pos	&	\pos	&	\pos	&	\pos	&	\pos	&	\pos	&	\pos	&	\pos	\\
{\Large\textbf{15-VMAF}}	&	\negat	&	\negat	&	\negat	&	\negat	&	\negat	&	\negat	&	\negat	&	\zer	&	\zer	&	\negat	&	\zer	&	\zer	&	\zer	&	\zer	&	\zer	&	\zer	&	\zer	&	\zer	&	\zer	&	\pos	&	\pos	&	\pos	&	\pos	&	\pos	&	\pos	&	\pos	&	\pos	\\
{\Large\textbf{16-PSNR-HVS}}	&	\negat	&	\negat	&	\negat	&	\negat	&	\negat	&	\negat	&	\negat	&	\negat	&	\negat	&	\negat	&	\negat	&	\negat	&	\zer	&	\negat	&	\zer	&	\zer	&	\zer	&	\zer	&	\zer	&	\zer	&	\zer	&	\zer	&	\pos	&	\pos	&	\pos	&	\pos	&	\pos	\\
{\Large\textbf{17-HAAR-PSI}}	&	\negat	&	\negat	&	\negat	&	\negat	&	\negat	&	\negat	&	\negat	&	\negat	&	\negat	&	\negat	&	\negat	&	\negat	&	\zer	&	\negat	&	\zer	&	\zer	&	\zer	&	\zer	&	\zer	&	\zer	&	\zer	&	\zer	&	\pos	&	\pos	&	\pos	&	\pos	&	\pos	\\
{\Large\textbf{18-FSIMc}}	&	\negat	&	\negat	&	\negat	&	\negat	&	\negat	&	\negat	&	\negat	&	\negat	&	\negat	&	\negat	&	\negat	&	\negat	&	\zer	&	\zer	&	\zer	&	\zer	&	\zer	&	\zer	&	\zer	&	\pos	&	\pos	&	\pos	&	\pos	&	\pos	&	\pos	&	\pos	&	\pos	\\
{\Large\textbf{19-LPIPS}}	&	\negat	&	\negat	&	\negat	&	\negat	&	\negat	&	\negat	&	\negat	&	\negat	&	\negat	&	\negat	&	\negat	&	\negat	&	\zer	&	\zer	&	\zer	&	\zer	&	\zer	&	\zer	&	\zer	&	\zer	&	\zer	&	\zer	&	\pos	&	\pos	&	\pos	&	\pos	&	\pos	\\
{\Large\textbf{20-FSIM}}	&	\negat	&	\negat	&	\negat	&	\negat	&	\negat	&	\negat	&	\negat	&	\negat	&	\negat	&	\negat	&	\negat	&	\negat	&	\negat	&	\negat	&	\negat	&	\zer	&	\zer	&	\negat	&	\zer	&	\zer	&	\zer	&	\zer	&	\pos	&	\pos	&	\pos	&	\pos	&	\pos	\\
{\Large\textbf{21-VSI}}	&	\negat	&	\negat	&	\negat	&	\negat	&	\negat	&	\negat	&	\negat	&	\negat	&	\negat	&	\negat	&	\negat	&	\negat	&	\negat	&	\negat	&	\negat	&	\zer	&	\zer	&	\negat	&	\zer	&	\zer	&	\zer	&	\zer	&	\pos	&	\pos	&	\pos	&	\pos	&	\pos	\\
{\Large\textbf{22-A-DISTS}}	&	\negat	&	\negat	&	\negat	&	\negat	&	\negat	&	\negat	&	\negat	&	\negat	&	\negat	&	\negat	&	\negat	&	\negat	&	\negat	&	\negat	&	\negat	&	\zer	&	\zer	&	\negat	&	\zer	&	\zer	&	\zer	&	\zer	&	\zer	&	\pos	&	\pos	&	\pos	&	\pos	\\
{\Large\textbf{23-PSNRY}}	&	\negat	&	\negat	&	\negat	&	\negat	&	\negat	&	\negat	&	\negat	&	\negat	&	\negat	&	\negat	&	\negat	&	\negat	&	\negat	&	\negat	&	\negat	&	\negat	&	\negat	&	\negat	&	\negat	&	\negat	&	\negat	&	\zer	&	\zer	&	\zer	&	\zer	&	\pos	&	\pos	\\
{\Large\textbf{24-DISTS}}	&	\negat	&	\negat	&	\negat	&	\negat	&	\negat	&	\negat	&	\negat	&	\negat	&	\negat	&	\negat	&	\negat	&	\negat	&	\negat	&	\negat	&	\negat	&	\negat	&	\negat	&	\negat	&	\negat	&	\negat	&	\negat	&	\negat	&	\zer	&	\zer	&	\zer	&	\pos	&	\pos	\\
{\Large\textbf{25-TOPIQ}}	&	\negat	&	\negat	&	\negat	&	\negat	&	\negat	&	\negat	&	\negat	&	\negat	&	\negat	&	\negat	&	\negat	&	\negat	&	\negat	&	\negat	&	\negat	&	\negat	&	\negat	&	\negat	&	\negat	&	\negat	&	\negat	&	\negat	&	\zer	&	\zer	&	\zer	&	\pos	&	\pos	\\
{\Large\textbf{26-FLIP}}	&	\negat	&	\negat	&	\negat	&	\negat	&	\negat	&	\negat	&	\negat	&	\negat	&	\negat	&	\negat	&	\negat	&	\negat	&	\negat	&	\negat	&	\negat	&	\negat	&	\negat	&	\negat	&	\negat	&	\negat	&	\negat	&	\negat	&	\negat	&	\negat	&	\negat	&	\zer	&	\pos	\\
{\Large\textbf{27-UQI}}	&	\negat	&	\negat	&	\negat	&	\negat	&	\negat	&	\negat	&	\negat	&	\negat	&	\negat	&	\negat	&	\negat	&	\negat	&	\negat	&	\negat	&	\negat	&	\negat	&	\negat	&	\negat	&	\negat	&	\negat	&	\negat	&	\negat	&	\negat	&	\negat	&	\negat	&	\negat	&	\zer	\\
\bottomrule
\end{tabular}
\end{adjustbox}
\caption{MRR test results for assessing the statistical significance between two objective quality metrics (sorted from the highest to the lowest SROCC). Higher performance for the metric in the column than the one in the row (black color), lower performance for the metric in the column than the one in the row (white color), equal performance between the metric in the row and column (gray color).}
\label{tbl_MRR_test}
\end{table}

\begin{table}[t]
\centering
\renewcommand{\arraystretch}{2}
\newcommand{\cellheader}[1]{\scriptsize\textbf{#1}}
\newcommand{\rot}[1]{\rotatebox{90}{\large\textbf{#1}}}
\newcommand{\negat}{\cellcolor{black!80}}
\newcommand{\zer}{\cellcolor{gray!30}}
\newcommand{\pos}{\cellcolor{white}}

\tiny
\begin{adjustbox}{max width=\columnwidth}
\begin{tabular}{
  >{\cellheader}l *{27}{>{\centering\arraybackslash}p{0.25cm}} 
}
\toprule
& \rot{01} & \rot{02} & \rot{03} & \rot{04} & \rot{05} & \rot{06} & \rot{07} & \rot{08} & \rot{09} & \rot{10} & \rot{11} & \rot{12} & \rot{13} & \rot{14} & \rot{15} & \rot{16} & \rot{17} & \rot{18} & \rot{19} & \rot{20} & \rot{21} & \rot{22} & \rot{23} & \rot{24} & \rot{25} & \rot{26} & \rot{27} \\
\midrule
{\Large \textbf{01-CVVDP}} & \zer & \pos & \pos & \pos & \pos & \pos & \pos & \pos & \pos & \pos & \pos & \pos & \pos & \pos & \pos & \pos & \pos & \pos & \pos & \pos & \pos & \pos & \pos & \pos & \pos & \pos & \pos \\
{\Large\textbf{02-IW-SSIM}} & \negat & \zer & \pos & \pos & \pos & \pos & \pos & \pos & \pos & \pos & \pos & \pos & \pos & \pos & \pos & \pos & \pos & \pos & \pos & \pos & \pos & \pos & \pos & \pos & \pos & \pos & \pos \\
{\Large \textbf{04-HDR-VDP-3 Q}} & \negat & \negat & \zer & \zer & \pos & \zer & \zer & \pos & \pos & \zer & \pos & \pos & \pos & \pos & \pos & \pos & \pos & \pos & \pos & \pos & \pos & \pos & \pos & \pos & \pos & \pos & \pos \\
{\Large\textbf{03-MS-SSIM}} & \negat & \negat & \zer & \zer & \zer & \zer & \pos & \pos & \pos & \pos & \pos & \pos & \pos & \pos & \pos & \pos & \pos & \pos & \pos & \pos & \pos & \pos & \pos & \pos & \pos & \pos & \pos \\
{\Large \textbf{05-NLPD}} & \negat & \negat & \zer & \negat & \zer & \zer & \zer & \pos & \zer & \zer & \pos & \pos & \pos & \pos & \pos & \pos & \pos & \pos & \pos & \pos & \pos & \pos & \pos & \pos & \pos & \pos & \pos \\
{\Large \textbf{07-VMAF-NEG}} & \negat & \negat & \negat & \zer & \zer & \zer & \zer & \zer & \zer & \zer & \zer & \pos & \pos & \pos & \pos & \pos & \pos & \pos & \pos & \pos & \pos & \pos & \pos & \pos & \pos & \pos & \pos \\
{\Large \textbf{09-SSIM}} & \negat & \negat & \negat & \negat & \zer & \zer & \zer & \zer & \zer & \zer & \zer & \zer & \pos & \pos & \pos & \pos & \pos & \pos & \pos & \pos & \pos & \pos & \pos & \pos & \pos & \pos & \pos \\
{\Large \textbf{10-PieAPP}}  & \negat & \negat & \negat & \zer & \zer & \zer & \zer & \zer & \zer & \zer & \zer & \zer & \pos & \pos & \pos & \zer & \pos & \pos & \pos & \pos & \pos & \pos & \pos & \pos & \pos & \pos & \pos \\
{\Large \textbf{06-HDR-VDP-2 Q}} & \negat & \negat & \zer & \zer & \zer & \zer & \zer & \zer & \zer & \zer & \zer & \pos & \pos & \pos & \pos & \pos & \pos & \pos & \pos & \pos & \pos & \pos & \pos & \pos & \pos & \pos & \pos \\
{\Large \textbf{08-SSIMULACRA1}} & \negat & \negat & \negat & \negat & \negat & \zer & \zer & \zer & \zer & \zer & \zer & \zer & \pos & \pos & \pos & \pos & \pos & \pos & \pos & \pos & \pos & \pos & \pos & \pos & \pos & \pos & \pos \\
{\Large \textbf{12-VIF}} & \negat & \negat & \negat & \negat & \negat & \negat & \negat & \zer & \zer & \zer & \zer & \zer & \pos & \pos & \zer & \zer & \zer & \zer & \pos & \pos & \pos & \pos & \pos & \pos & \pos & \pos & \pos \\
{\Large\textbf{11-SSIMULACRA2}} & \negat & \negat & \negat & \negat & \negat & \zer & \zer & \zer & \zer & \zer & \zer & \zer & \pos & \pos & \pos & \zer & \pos & \zer & \pos & \pos & \pos & \pos & \pos & \pos & \pos & \pos & \pos \\
{\Large\textbf{17-BUTTERAUGLI}} & \negat & \negat & \negat & \negat & \negat & \negat & \negat & \negat & \negat & \negat & \negat & \zer & \zer & \zer & \zer & \zer & \zer & \zer & \zer & \zer & \pos & \pos & \pos & \pos & \pos & \pos & \pos \\
{\Large\textbf{15-GMSD}}  & \negat & \negat & \negat & \negat & \negat & \negat & \negat & \negat & \negat & \negat & \negat & \zer & \negat & \negat & \zer & \zer & \zer & \zer & \zer & \pos & \pos & \pos & \pos & \pos & \pos & \pos & \pos \\
{\Large\textbf{16-VMAF}} & \negat & \negat & \negat & \negat & \negat & \negat & \negat & \negat & \negat & \zer & \zer & \zer & \zer & \zer & \zer & \zer & \zer & \zer & \negat & \pos & \pos & \pos & \pos & \pos & \pos & \pos & \pos \\
{\Large\textbf{14-PSNR-HVS}} & \negat & \negat & \negat & \negat & \negat & \negat & \negat & \negat & \negat & \negat & \negat & \negat & \zer & \zer & \pos & \zer & \zer & \zer & \zer & \zer & \pos & \pos & \pos & \pos & \pos & \pos & \pos \\
{\Large\textbf{13-HAAR-PSI}} & \negat & \negat & \negat & \negat & \negat & \negat & \negat & \negat & \negat & \negat & \negat & \negat & \zer & \zer & \pos & \zer & \zer & \zer & \zer & \zer & \pos & \pos & \pos & \pos & \pos & \pos & \pos \\
{\Large\textbf{20-FSIMc}} & \negat & \negat & \negat & \negat & \negat & \negat & \negat & \negat & \negat & \negat & \negat & \negat & \zer & \zer & \negat & \negat & \zer & \zer & \zer & \zer & \pos & \pos & \pos & \pos & \zer & \pos & \pos \\
{\Large\textbf{18-LPIPS}} & \negat & \negat & \negat & \negat & \negat & \negat & \negat & \negat & \negat & \negat & \zer & \zer & \zer & \zer & \zer & \zer & \zer & \zer & \zer & \zer & \pos & \pos & \pos & \pos & \pos & \pos & \pos \\
{\Large\textbf{19-FSIM}}  & \negat & \negat & \negat & \negat & \negat & \negat & \negat & \negat & \negat & \negat & \negat & \negat & \zer & \zer & \zer & \pos & \zer & \zer & \zer & \zer & \pos & \pos & \pos & \pos & \zer & \pos & \pos \\
{\Large\textbf{22-VSI}} & \negat & \negat & \negat & \negat & \negat & \negat & \negat & \negat & \negat & \negat & \negat & \negat & \negat & \negat & \negat & \negat & \negat & \negat & \negat & \negat & \zer & \zer & \pos & \zer & \zer & \pos & \pos \\
{\Large\textbf{21-A-DISTS}} & \negat & \negat & \negat & \negat & \negat & \negat & \negat & \negat & \negat & \negat & \negat & \negat & \negat & \negat & \negat & \negat & \negat & \negat & \negat & \negat & \zer & \zer & \pos & \zer & \zer & \pos & \pos \\
{\Large\textbf{25-PSNRY}} & \negat & \negat & \negat & \negat & \negat & \negat & \negat & \negat & \negat & \negat & \negat & \negat & \negat & \negat & \negat & \negat & \negat & \negat & \zer & \zer & \zer & \zer & \zer & \zer & \zer & \pos & \pos \\
{\Large\textbf{23-DISTS}} & \negat & \negat & \negat & \negat & \negat & \negat & \negat & \negat & \negat & \negat & \negat & \negat & \negat & \negat & \negat & \negat & \negat & \negat & \negat & \negat & \negat & \negat & \zer & \zer & \zer & \pos & \pos \\
{\Large\textbf{24-TOPIQ}} & \negat & \negat & \negat & \negat & \negat & \negat & \negat & \negat & \negat & \negat & \negat & \negat & \negat & \negat & \negat & \negat & \negat & \negat & \negat & \negat & \zer & \zer & \zer & \zer & \zer & \pos & \pos \\
{\Large\textbf{26-FLIP}} & \negat & \negat & \negat & \negat & \negat & \negat & \negat & \negat & \negat & \negat & \negat & \negat & \negat & \negat & \negat & \negat & \negat & \negat & \negat & \negat & \negat & \negat & \negat & \negat & \negat & \zer & \pos \\
{\Large\textbf{27-UQI}}  & \negat & \negat & \negat & \negat & \negat & \negat & \negat & \negat & \negat & \negat & \negat & \negat & \negat & \negat & \negat & \negat & \negat & \negat & \negat & \negat & \negat & \negat & \negat & \negat & \negat & \negat & \zer \\
 \bottomrule
\end{tabular}
\end{adjustbox}
\caption{Wilcoxon Signed-Rank test results between two metrics (sorted from the highest to the lowest SROCC). Higher performance for the metric in the column than the one in the row (black color), lower performance for the metric in the column than the one in the row (white color), equal performance between the metric in the row and column (gray color).}
\label{tbl-Wincox}
\end{table}

\subsection{Studying the Cropping Impact}\label{lbl-resolution}
The purpose of this experiment is to investigate the impact of the cropping on the performance of objective quality assessment metrics by applying the proposed statistical tests described in Section~\ref{sec:stat-test}. While the overall evaluation of metrics in the previous section is conducted using the entire full-resolution images (in a practical scenario the crop locations are not known), the subjective quality scores were obtained using cropped smaller resolution image patches. This setup raises the question of whether the mismatch in resolution or the area covered between subjective and objective assessments introduces any bias or inconsistency in the evaluation process. The first step was to jointly transform the objective quality scores using a logistic function, as explained in Section \ref{sec:transformation}, using the metric values from both full-resolution and cropped images.

\begin{figure*}[!t]
    \centering
    \includegraphics[width=1\linewidth]{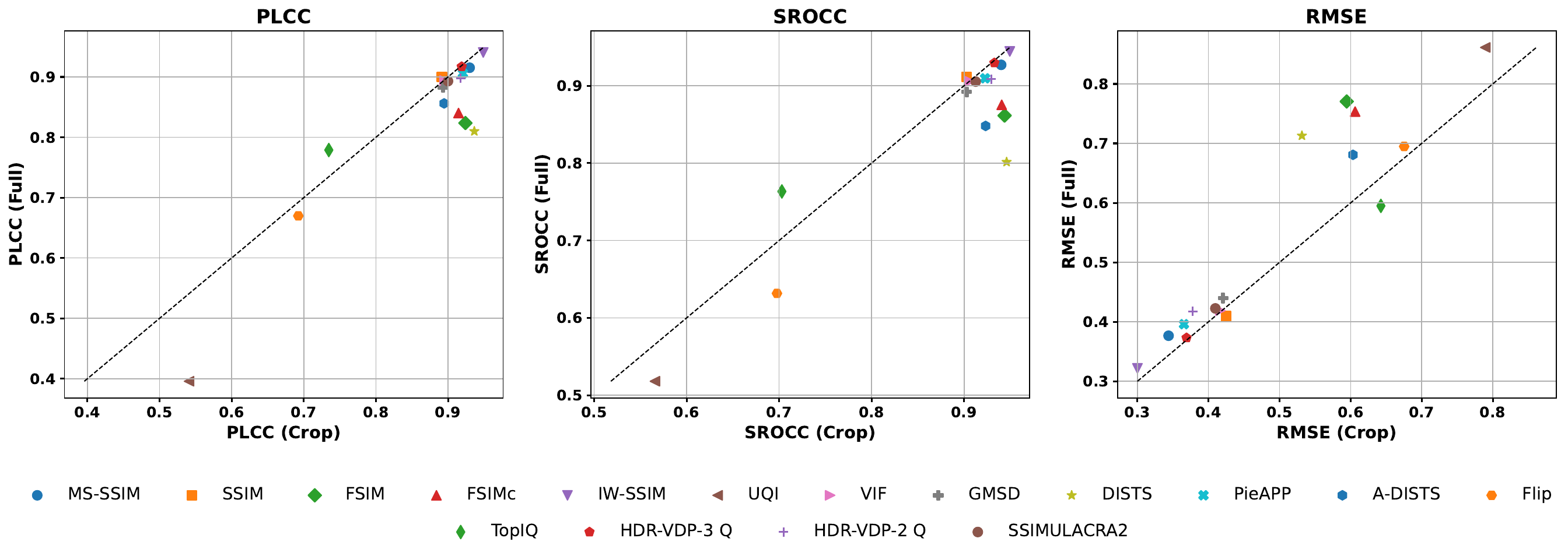}
    \caption{Performance comparison of objective quality metrics applied to cropped versus full-resolution images. Only metrics showing statistically significant differences are included.}
    \label{fig:spatial_resolution_performance}
\end{figure*}

\begin{figure}[!t]
    \centering
    \begin{subfigure}[b]{0.495\linewidth}
        \centering
        \includegraphics[scale=0.46]{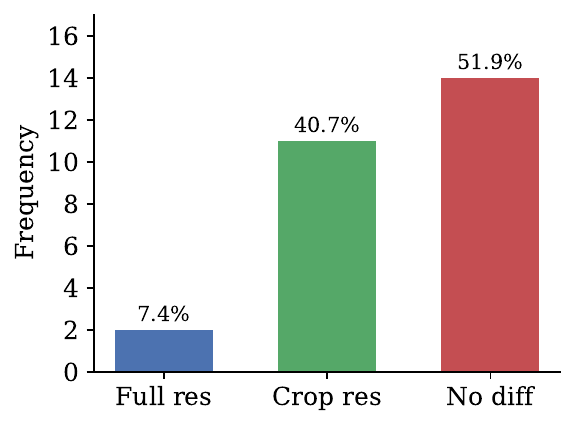}
        \caption{Meng–Rosenthal–Rubin (MRR) test}
    \end{subfigure}
    \begin{subfigure}[b]{0.495\linewidth}
        \centering
        \includegraphics[scale=0.46]{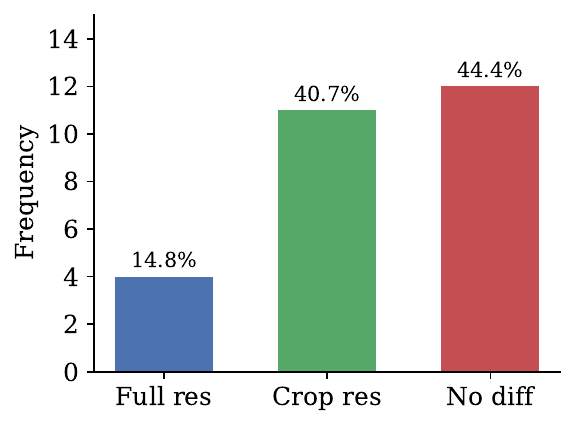}
        \caption{Pairwise Wilcoxon Signed-Rank test }
    \end{subfigure}
    \caption{Statistical test outcome of the impact of cropping on the performance of objective image quality metrics.}
    \label{fig:SI_on_resolution}
\end{figure}

The Meng–Rosenthal–Rubin (MRR) test for SROCC and the Pairwise Wilcoxon Signed-Rank test as described in Section~\ref{sec:stat-test} were applied to assess statistically significance difference. The results, shown in Figure~\ref{fig:SI_on_resolution}, indicate that both tests yield similar outcomes, with approximately 50\% of cases showing no significant difference between the performance of the metrics on cropped and full-resolution images. As expected, metrics using cropped images are better since in this case it is consistent with the subjective assessment procedure.

Additionally, the performance of metrics computed on cropped images versus full-resolution images is compared, using the most common evaluation PLCC, SROCC, and RMSE criteria for those metrics that show statistical significance difference. This difference can be attributed to two main factors: first, objective quality metrics are sensitive to the spatial resolution of the input; second, cropping the images may lead to variations in subjective quality scores when comparing full-resolution versus cropped versions shown to participants.

The experimental results are presented in Figure~\ref{fig:spatial_resolution_performance} for the metrics that are statistically difference for both MRR and Wilcoxon Signed-Rank test. As shown, the performance differences using the selected evaluation criteria are generally not very high. The maximum observed difference is observed for the UQI, DISTS, A-DISTS, FSIM, FSIMc, TOPIQ, which are quality metrics that do not perform very well (for PLCC, SROC and RMSE) in the analysis presented in Section \ref{lbl-performance}. For other metrics, such as IW-SSIM, HDR-VDP-2/3, the performance difference is rather small which can be observed in Figure~\ref{fig:spatial_resolution_performance} since the the corresponding point is very close to the identity line.
\section{Conclusions and Future Work}
As compression algorithms push toward visually lossless performance, the need for accurate and sensitive image quality assessment metrics becomes increasingly important. This study provides a detailed evaluation of existing image quality metrics within the high quality to mathematically lossless compression regime using the very recent and reliable JPEG AIC-3 dataset. Moreover, new evaluation measures were proposed and novel statistical tests were introduced. Our findings demonstrate that many widely used metrics, while effective at lower quality levels, struggle to distinguish subtle perceptual differences that are still meaningful to human observers in high-quality images. The experimental results show that CVVDP consistently has higher performance across multiple evaluation criteria and statistical tests, demonstrating strong alignment with human perceptual judgments in the near-lossless range. IW-SSIM emerged as the next best-performing metric, offering robust perceptual relevance, albeit with slightly lower performance than CVVDP. For future work, it is suggested to extend this study by employing a more diverse image dataset, obtained through a larger subjective study that includes a wide range of image sources and degradation types. Additionally, new quality metrics, expected to be proposed in response to the JPEG AIC-4 Call for Proposals, will be evaluated using this work as a benchmark, providing a solid foundation for comparison and further validation.

\bibliographystyle{ieeetr}
\bibliography{refs}

\begin{IEEEbiography}[{\includegraphics[width=1in,height=1.25in,clip,keepaspectratio]{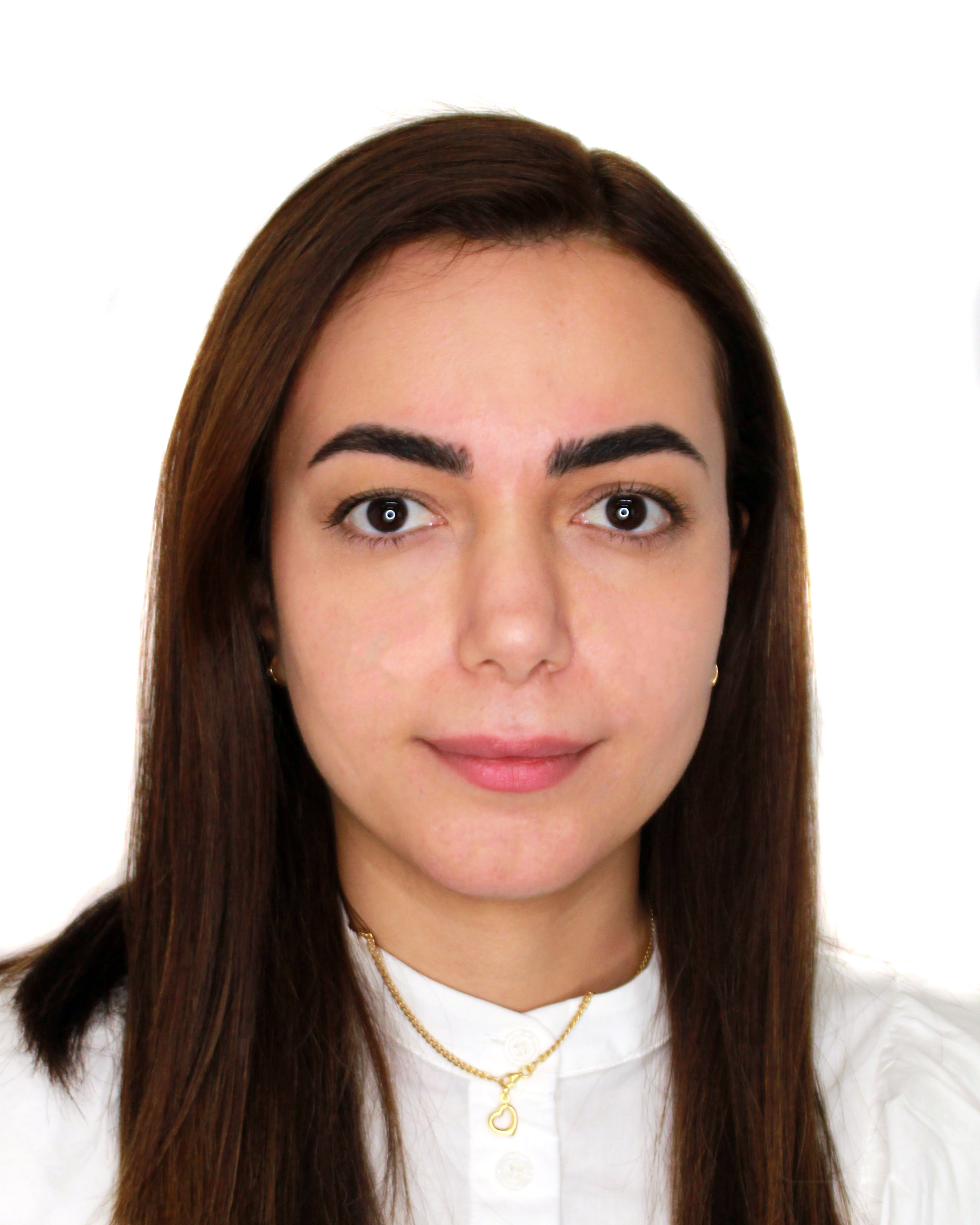}}]{Shima Mohammadi} (Graduate Student Member, IEEE) 
received the M.Sc. degree in Electrical and Computer Engineering from the University of Tehran, Tehran, Iran, in 2020. She is currently a Ph.D. student with the Department of Electrical and Computer Engineering, Instituto Superior Técnico (IST), Universidade Técnica de Lisboa, Lisbon, Portugal, and a member of the Instituto de Telecomunicações. Her research interests include visual coding, quality assessment, and machine learning applications.
\end{IEEEbiography}
\vskip 0pt plus -1fil

\begin{IEEEbiography}[{\includegraphics[width=1in,height=2in,clip,keepaspectratio]{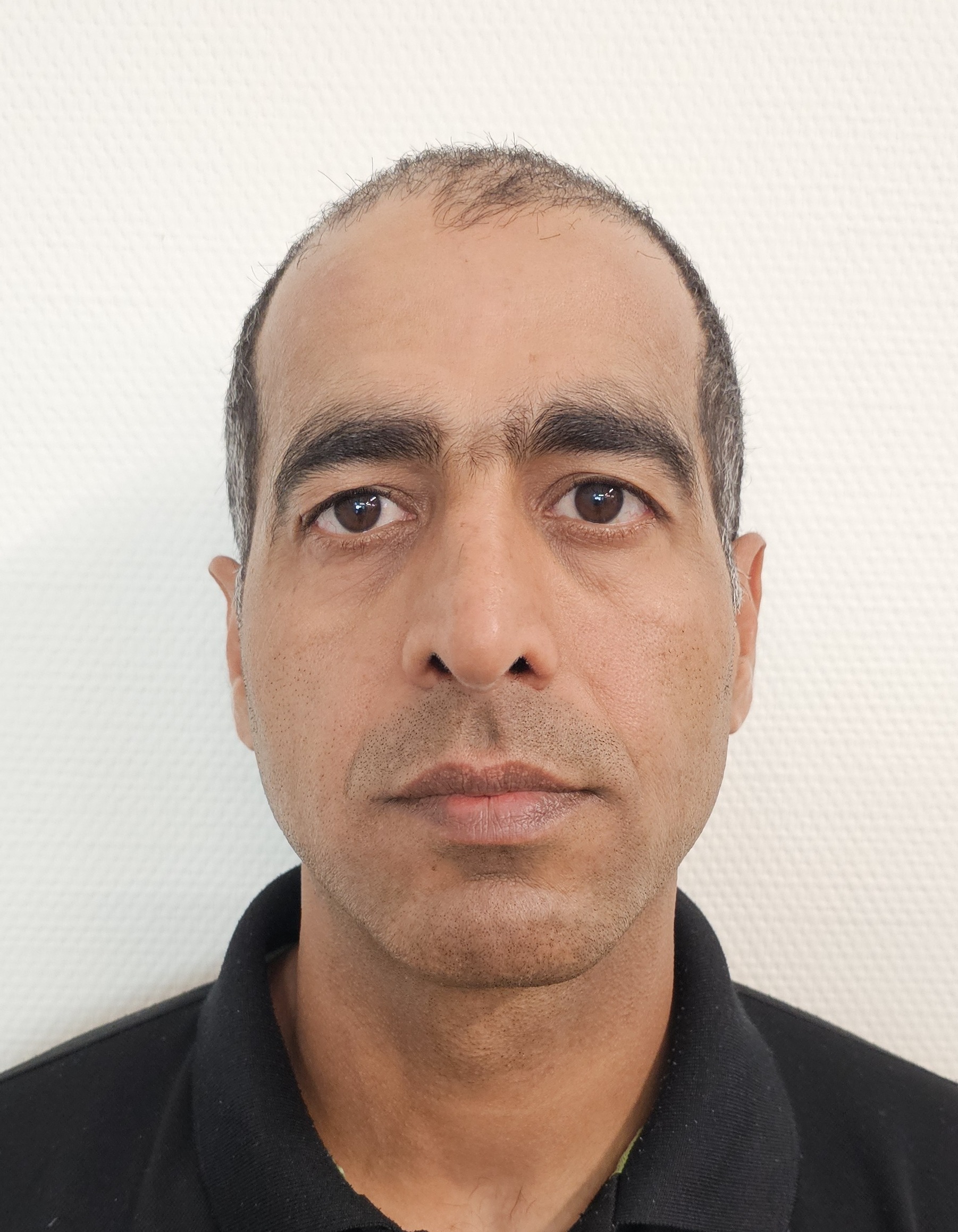}}]{Mohsen Jenadeleh} (Member, IEEE) Mohsen Jenadeleh (Member, IEEE) received his Dr.rer.nat. in 2019 from the Department of Computer and Information Science at the University of Konstanz, Germany. He continues to pursue his research at the same institution as a postdoctoral researcher. Currently, he is leading a research project titled ``JND-based perceptual video quality analysis and modelling'' funded by the German Research Foundation (DFG) -- Project ID 496858717 through a grant for Temporary Positions for Principal Investigators. His research interests span image and video processing, perceptual quality assessment of visual media, machine learning, deep learning, and crowdsourcing. He is involved in activities organized by standardization committees such as JPEG-AIC and VQEG.
\end{IEEEbiography}
\vskip 0pt plus -1fil

\begin{IEEEbiography}[{\includegraphics[width=1in,height=1.25in,clip,keepaspectratio]{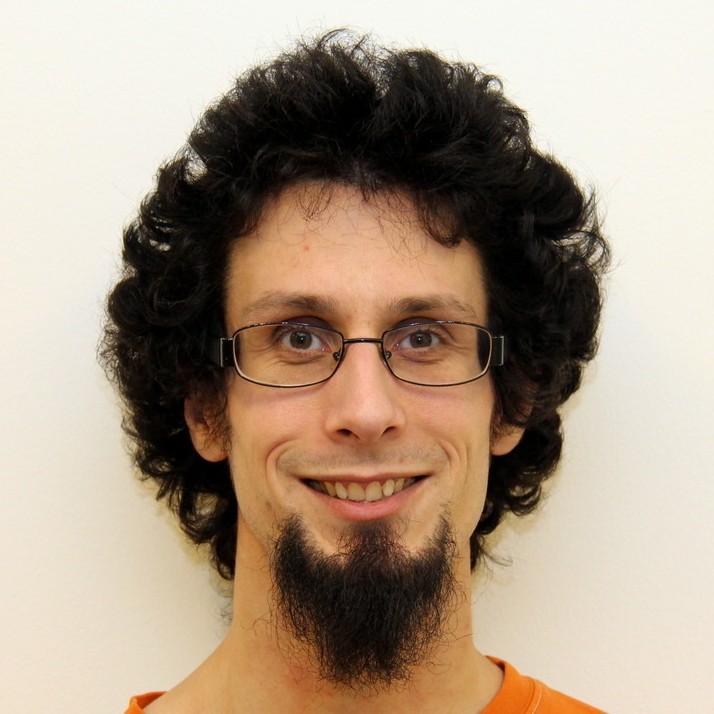}}]{Jon Sneyers} received the M.Sc. and Ph.D. degrees in computer science from the KU Leuven, Belgium, in 2004 and 2008, respectively. Since 2016 he is researching image processing and image compression at Cloudinary, where currently he is a Staff Researcher. He is the co-creator of the FLIF, FUIF and JPEG XL image formats, and is project lead of the ISO/IEC 18181 standard (JPEG XL).

\end{IEEEbiography}

\begin{IEEEbiography}[{\includegraphics[width=1in,height=2.5in,clip,keepaspectratio]{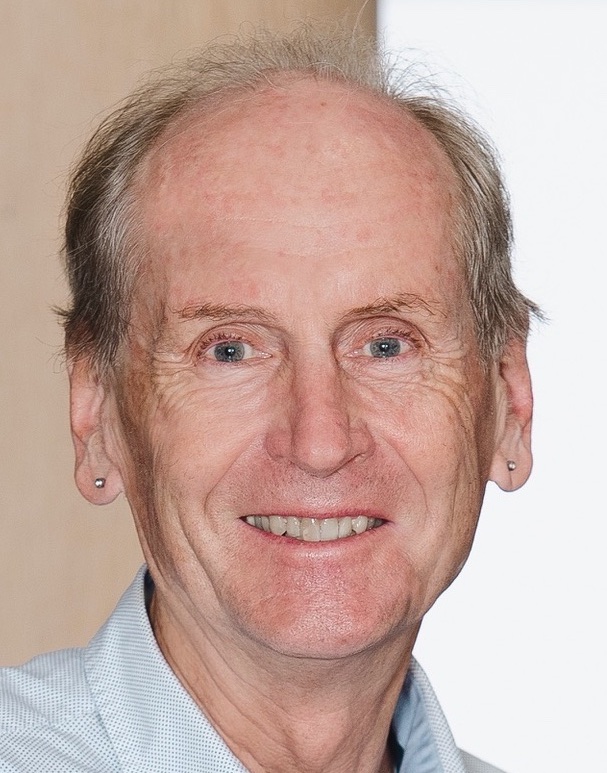}}]{Dietmar Saupe} received the Dr.\ rer.\ nat.\ and Habilitation degrees in mathematics from the University of Bremen, Germany, in 1982 and 1993, respectively. 
From 1985 to 1993, he was an Assistant Professor with the Departments of Mathematics, first at the University of California, Santa Cruz, USA, and then at the University of Bremen. From 1993 to 1998, he was a Professor of Computer Science with the University of Freiburg, Germany, the University of Leipzig, Germany, until 2002, and since then, the University of Konstanz, Germany. He is the coauthor of the book Chaos and Fractals (Springer-Verlag, 1992), which won the Association of American Publishers Award for Best Mathematics Book of the Year, the book The Science of Fractal Images (Springer-Verlag, 1988), and well over 100 research articles. His research interests include image and video processing, computer graphics, scientific visualisation, dynamical systems, and sport informatics.
\end{IEEEbiography}
\vskip 0pt plus -1fil

\begin{IEEEbiography}[{\includegraphics[width=1in,height=1.25in,clip,keepaspectratio]{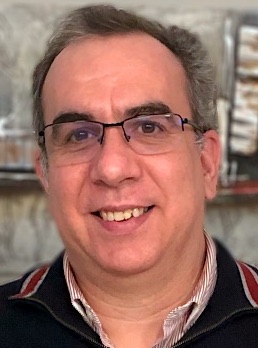}}]{Joao Ascenso} (Senior Member, IEEE)
received the E.E., M.Sc., and Ph.D. degrees in electrical and computer engineering from the Instituto Superior Técnico (IST), Universidade Técnica de Lisboa, Lisbon, Portugal, in 1999, 2003, and 2010, respectively. He is currently an Associate Professor with the Department of Electrical and Computer Engineering, IST, and a member of the Instituto de Telecomunicações. He has published more than 150 papers in international conferences and journals. His current research interests include visual coding, quality assessment, coding and processing of 3D visual representations, coding for machines, super-resolution, denoising among others. He was an associate editor of IEEE Transactions on Image Processing, IEEE Signal Processing Letters and IEEE Transactions on Multimedia and is currently senior area editor of IEEE Transactions on Circuits and Systems for Video Technology. He received three Best Paper Awards at PCS 2015, ICME 2020, and MMSP 2024 and has served as Technical Program Chair and in other organizing committees of major international conferences, including IEEE ICIP, PCS, EUVIP, ICME, MMSP and ISM. 
\end{IEEEbiography}

\appendix
\begin{figure*}[!htbp]

\centerline{
\begin{tabular}{@{}c@{}}
    {\includegraphics[scale=0.28]{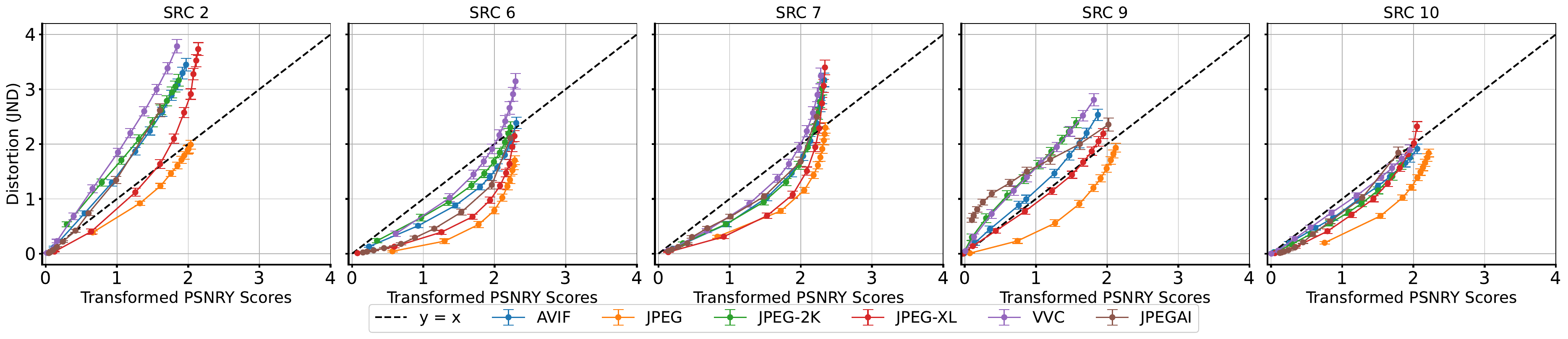}} \\
  \end{tabular}
}

\centerline{
\begin{tabular}{@{}c@{}}
    {\includegraphics[scale=0.28]{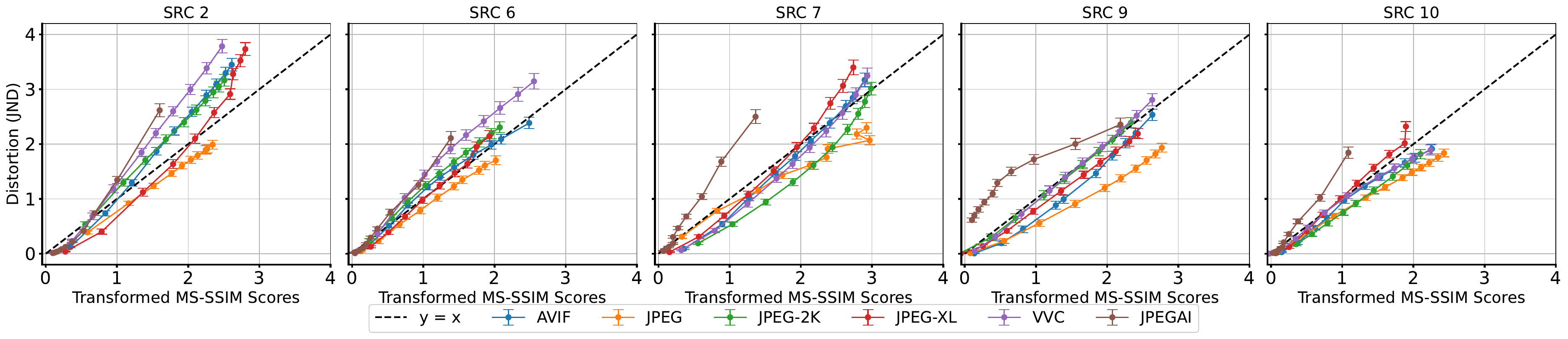}} \\
  \end{tabular}}

\centerline{
\begin{tabular}{@{}c@{}}
    {\includegraphics[scale=0.28]{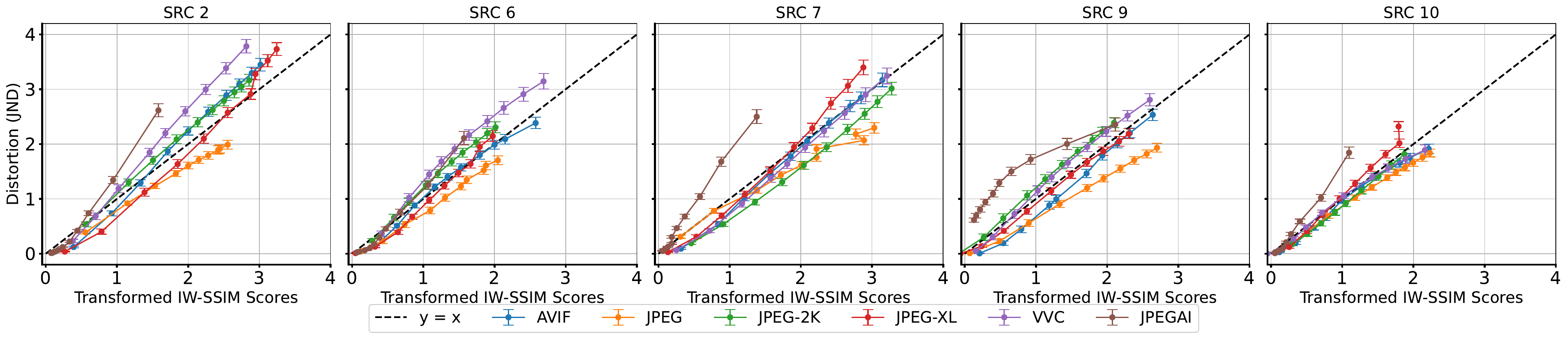}} \\
  \end{tabular}
}

\centerline{
\begin{tabular}{@{}c@{}}
    {\includegraphics[scale=0.28]{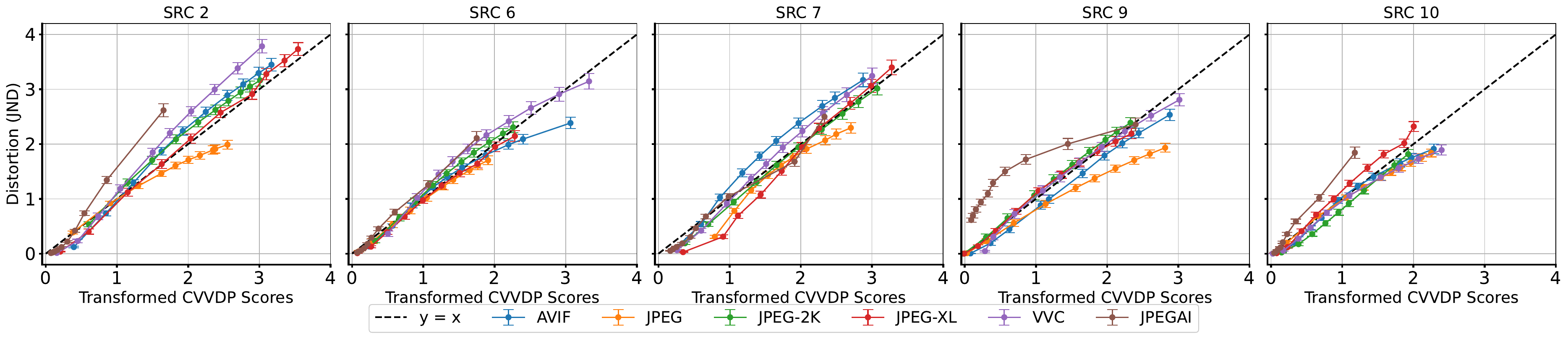}} \\
  \end{tabular}
}

\centerline{
\begin{tabular}{@{}c@{}}
    {\includegraphics[scale=0.28]{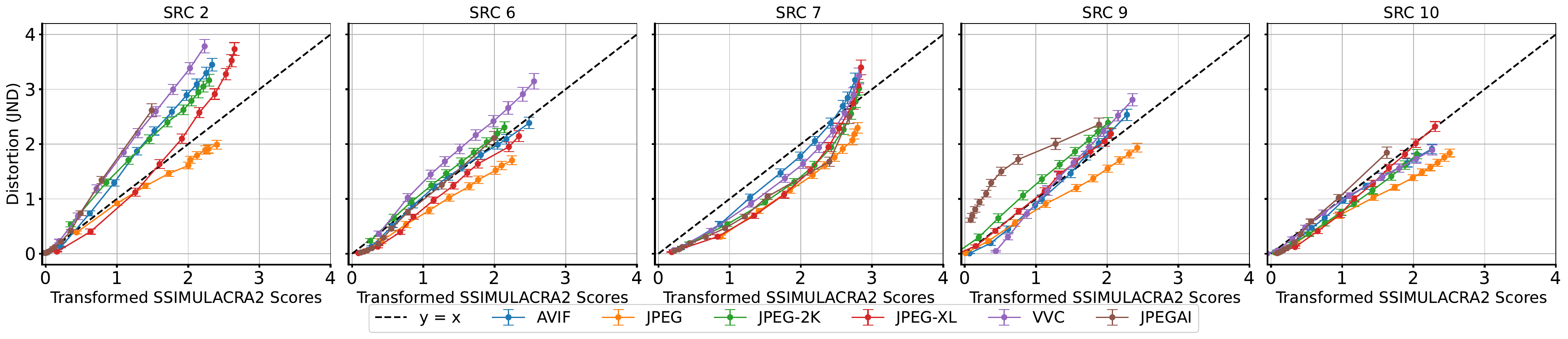}} \\
  \end{tabular}
}

\centerline{
\begin{tabular}{@{}c@{}}
    {\includegraphics[scale=0.28]{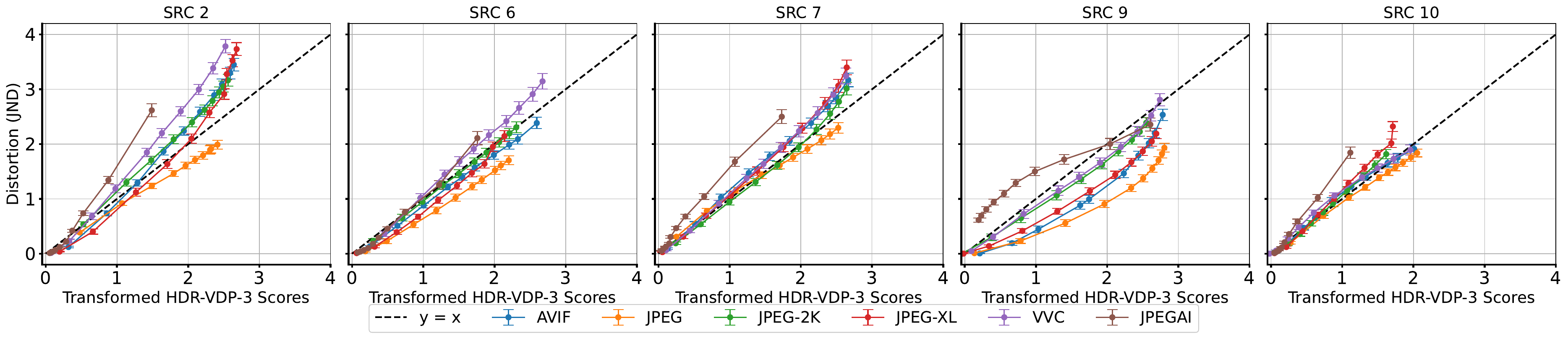}} \\
  \end{tabular}
}
\caption{Subjective scores vs transformed objective scores}  
\label{Fig_model_vs_pref}
\end{figure*}

\vfill

\EOD

\end{document}